\documentclass[openacc]{rstransa}

\DeclareMathOperator*{\argmax}{arg\,max}
\DeclareMathOperator{\sgn}{sgn}

\usepackage{bm}
\usepackage{braket}
\usepackage{overpic}
\usepackage{algorithm,algorithmic}
\usepackage[numbers,sort&compress]{natbib}

\begin{document}

\title{Greedy parameter optimization for diabatic quantum annealing}

\author{
Tadashi Kadowaki$^{1}$ and Hidetoshi Nishimori$^{2,3,4}$}

\address{$^{1}$DENSO CORPORATION, Kounan, Minato-ku, Tokyo 108-0075, Japan\\
$^{2}$Institute of Innovative Research, Tokyo Institute of Technology, Yokohama, Kanagawa 226-8503, Japan\\
$^{3}$Graduate School of Information Sciences, Tohoku University, Sendai, Miyagi 980-8579, Japan\\
$^{4}$RIKEN Interdisciplinary Theoretical and Mathematical Sciences Program (iTHEMS), Wako, Saitama 351-0198, Japan}

\subject{quantum physics, quantum information, computer science}

\keywords{quantum annealing, counterdiabatic driving, greedy optimization}

\corres{Tadashi Kadowaki\\
\email{tadashi.kadowaki.j3m@jp.denso.com}}

\begin{abstract}
A shorter processing time is desirable for quantum computation to minimize the effects of noise.  We propose a simple procedure to variationally determine a set of parameters in the transverse-field Ising model for quantum annealing appended with a field along the $y$ axis. The method consists of greedy optimization of the signs of coefficients of the $y$-field term based on the outputs of short annealing processes. We test the idea in the ferromagnetic system with all-to-all couplings and spin-glass problems, and find that the method outperforms the traditional form of quantum annealing and simulated annealing in terms of the success probability and the time to solution, in particular in the case of shorter annealing times, achieving the goal of improved performance while avoiding noise. The non-stoquastic $\sigma^y$ term can be eliminated by a rotation in the spin space, resulting in a non-trivial diabatic control of the coefficients in the stoquastic transverse-field Ising model, which may be feasible for experimental realization.
\end{abstract}


\begin{fmtext}
\end{fmtext}

\maketitle

\section{Introduction}

Quantum optimization algorithm is an active field of research, including quantum annealing (QA) \cite{Kadowaki1998, Kadowaki1998a, Brooke1999, Santoro2002, Santoro2006, Das2008, Morita2008, Tanaka2017}, adiabatic quantum computing \cite{Farhi2001}, and quantum approximate optimization algorithm (QAOA) \cite{Farhi2014}.   If QA operates in the adiabatic regime, it often encounters a difficulty of exponential computation time coming from exponential closing of the energy gap due to a first-order phase transition, examples of which are found in Refs.~\cite{Jorg2008,Young2010,Jorg2010}.  A number of approaches have been tried to mitigate this problem, including non-stoquastic Hamiltonians \cite{Seki2012,Seki2015,Seoane2012,Nishimori2017}, inhomogeneous field driving \cite{Susa2018,Susa2018a,Mohseni2018,Adame2020,Rams2016,Gomez-Ruiz2019}, reverse annealing and pausing \cite{Ohkuwa2018,Yamashiro2019,Passarelli2020,Venturelli2019,Chancellor2020,Marshall2019,Ikeda2019,Grant2020,Rocutto2021,Arai2021,Golden2021}, and counterdiabatic driving \cite{Demirplak2003,Demirplak2005,Berry2009,Chen2010,Chen2011,Takahashi2013,Jarzynski2013,Torrontegui2012,Campo2013,Takahashi2017,Sels2017,Ozguler2018,Claeys2019,Hartmann2019,Hatomura2021}.

The last method of counterdiabatic driving is particularly appealing because it allows us to reduce the computation time as compared to the adiabatic process, yet keeping the success probability large.  The reduction of computation time is important not just theoretically but also experimentally from the viewpoint of avoiding the effects of noise in quantum devices.  One of the drawbacks of the idea of counterdiabatic driving is the excessive complexity of the additional counterdiabatic term, see e.g. Ref.~\cite{Takahashi2013}, which generally makes it impossible to be implemented experimentally.  An interesting development to avoid this problem is the variational approach to optimize the coefficients of approximate counterdiabatic terms composed of manageable operators \cite{Sels2017,Takahashi2017,Ozguler2018,Prielinger2021,Hatomura2021,Imoto2021}. In the present paper, we further simplify this strategy and propose to determine just the signs of coefficients of counterdiabatic terms composed of $y$ components of the Pauli matrix. Since the determination of signs is processed in a greedy way variationally, we call the method quantum greedy optimization (QGO).  This approach is appealing because of its simplicity and yet a significantly improved success probability in comparison with the original QA and the classical method of simulated annealing.

This paper is organized as follows. Section \ref{sec:formulation} defines the model system and explains our strategy to solve optimization problems. The performance of the method is analyzed numerically in section~\ref{sec:numerical} for the mean-field, ferromagnetic, and spin glass problems, and the results are compared with simulated annealing (SA) \cite{Kirkpatrick1983} and the original QA. Also discussed are a further simplification of the algorithm and a possible improvement based on an oracle setting. The final section \ref{sec:summary} summarizes and discusses the results.

\section{Formulation}
\label{sec:formulation}

Let us define the following time-dependent Hamiltonian
\begin{align}
    \mathcal{H} = A(t) \mathcal{H}^z + B(t) \mathcal{H}^x + \sum_{i=1}^N C_i(t) \mathcal{H}_i^y,
    \label{eq:Hamiltonian_full}
\end{align}
where $\mathcal{H}^z$, $\mathcal{H}^x$, and $\mathcal{H}_i^y$ are the problem (Ising) Hamiltonian, the transverse field term (to be called the $x$-field), and the $y$-field term, respectively,
\begin{align}
    \mathcal{H}^z = - \sum_{i<j} J_{ij} \sigma_i^z \sigma_j^z,~\mathcal{H}^x = - \sum_{i=1}^N \sigma_i^x, ~\mathcal{H}_i^y = - \sigma_i^y.
\end{align}
The symbols $\sigma_i^x, \sigma_i^y$, and $\sigma_i^z$ denote the components of the Pauli matrix at site (qubit) $i$, and $J_{ij}$ is for the interaction between sites $i$ and $j$. The problem size is $N$.  The term $\mathcal{H}_i^y$ renders the system non-stoquastic \cite{Bravyi2009} and was introduced in Ref.~\cite{Sels2017} to approximately realize counterdiabatic driving. See also Refs.~ \cite{Takahashi2017,Hartmann2019,Prielinger2021,Passarelli2020a}.  The time dependent coefficients $A(t), B(t)$, and $C_i(t)$ should satisfy the initial ($t=0$) and final $(t=\tau)$ conditions,
\begin{align}
 A(0)=C_i(0)=0, B(0)\ne 0,~ A(\tau)>0, B(\tau)=C_i(\tau)=0    
\end{align}
such that the system starts with the simple transverse field term $\mathcal{H}^x$ and ends with the problem Hamiltonian $\mathcal{H}^z$. The $y$-field is applied only in the middle of annealing.  We arbitrarily choose the following functions for these coefficients for simplicity though other (similar) functions can be considered \cite{Sels2017},
\begin{align}
    A(t) = \frac{at}{\tau},~B(t) = b\Big(1-\frac{t}{\tau}\Big),~C_i(t) = c_i \,\sin^2 \Big(\frac{\pi t}{\tau}\Big)
\end{align}
with $a=1$ to set the overall energy scale. The annealing time $\tau$ can be chosen arbitrarily and we test the cases of $\tau=1$ and $\tau=5$ in the following.

Our strategy is to optimize $b$ and $c_i$ under a variational principle to optimize appropriate measures.  We simplify the process by fixing the amplitude of $c_i$ to a value found optimal in the mean-field case (to be described below) and choosing only the sign of $c_i$ in a greedy way. This facilitates the process considerably and yet will turn out to lead to significant performance improvements.

Our variational optimization is carried out by minimization of the expectation value of the final energy $E = \braket{ \psi(t=\tau) | \mathcal{H}^z | \psi(t=\tau) }$, where $\psi(t)$ is the wave function at time $t$, as well as by maximization of the ground-state probability, or the fidelity, $P_{\rm gs} = |\braket{\psi(t=\tau) | \psi_{\rm gs}}|^2$, where $|\psi_{\rm gs}\rangle$ is the true ground state of the Ising Hamiltonian. Since we do not know the true ground state in a generic problem, we refer to the latter measure as an oracle.

The additional $y$-field term may not be straightforward to be implemented experimentally. Nevertheless, by a simple rotation of the spin axes \cite{Sels2017,Prielinger2021},
\begin{align}
    U_g(t) = \exp \left(\frac{i}{2} \sum_i \theta_i(t) \sigma_i^z \right) 
\end{align}
with $\theta_i(t) = \arctan (C_i(t) / B(t))$, we can transform the Hamiltonian into a transverse-field Ising model with non-trivial time dependence of coefficients,
\begin{align}
    \mathcal{H}_{\rm eff} =
        A(t) \mathcal{H}^z - \frac{1}{2}\sum_i   \frac{d\theta_i(t) }{dt} \,\sigma_i^z
        - \sum_i \sqrt{B(t)^2 + C_i(t)^2}\,\sigma_i^x  .
        \label{eq:Heff}
\end{align}
which may be easier to realize experimentally than the original Hamiltonian of Eq. (\ref{eq:Hamiltonian_full}).
An example of the behavior of the coefficients in Eq.~(\ref{eq:Heff}) is shown in Appendix \ref{appendix:coefficients}.

\section{Numerical results}
\label{sec:numerical}

In this section, we present numerical results for the mean-field, ferromagnetic, and random (spin glass) systems.  In addition, we discuss an additional algorithm which shortcuts the process of iterative determination of the signs of coefficients.

\subsection{Mean-field theory}
\label{subsec:meanfield}

As a preliminary to the next section of a ferromagnetic system, we first analyze the properties of the mean-field theory with the Hamiltonian (see Ref.~\cite{Takahashi2017} for a related idea)
\begin{equation}
	\mathcal{H} = -t \braket{\sigma^z} \sigma^z - b (1-t) \sigma^x - c \sin^2(\pi t) \sigma^y ,
\end{equation}
where $\braket{\sigma^z}$ is the magnetization of the system $\braket{\psi (t)| \sigma^z |\psi(t)}$. We set $\tau = 1$ in this section. The state vector can be expressed as a qubit $\ket{\psi(t)} = \alpha \ket{0} + \beta \ket{1}$, where $\ket{0} = (1,0)^T$ and $\ket{1} = (0, 1)^T$, and the Schr\"odinger equation reads
\begin{equation}
    \frac{d}{dt} \left( \begin{array}{c} \alpha \\ \beta \end{array} \right) =
    \left( \begin{array}{cc} i t \braket{\sigma^z} & ib(1-t) + c \sin^2(\pi t) \\ ib(1-t) - c \sin^2(\pi t) & -i t \braket{\sigma^z} \end{array} \right) \left( \begin{array}{c} \alpha \\ \beta \end{array} \right) ,
\end{equation}
where $\braket{\sigma^z}$ is $|\alpha|^2-|\beta|^2$.
As the ground state of the final Hamiltonian is doubly degenerate, we choose the spin-up  state $\braket{\sigma^z} > 0$ as the desired ground state. Two measures of success as defined in the previous section, the fidelity $P_{\rm gs}=(\braket{\sigma^z} +1)/2$ and the energy $E=-\braket{\sigma^z}^2$, can be expressed by the magnetization $\braket{\sigma^z}$ in the mean-field theory, which is not the case generally.

\begin{figure}[thb]
    \centering
    \includegraphics[width=80mm]{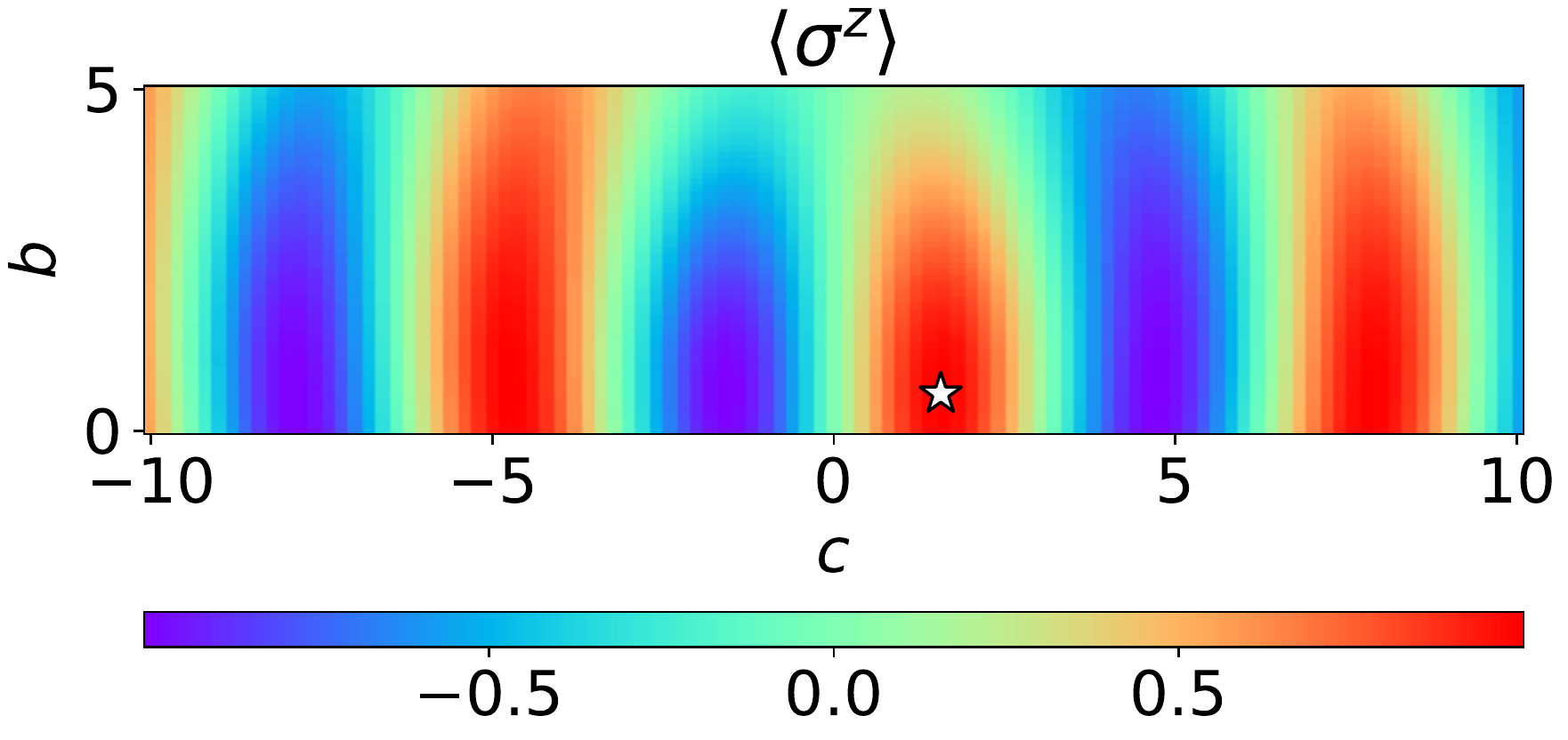}
    \caption{
        \label{fig1}
        Parameter dependence of the magnetization $\braket{\sigma^z}$ for the mean-field theory. The optimal point $(b_{\rm opt}^{\rm mf}=0.539, c_{\rm opt}^{\rm mf}=1.565)$ is marked in an open asterisk.
    }
\end{figure}

Figure~\ref{fig1} shows the dependence of magnetization $\braket{\sigma^z(t = 1)}$ on $b$ and $c$.
In this landscape of the parameter space, we find the optimal values to maximize  $\braket{\sigma^z}$ by the classical optimizer, the Broyden-Fletcher-Goldfarb-Shanno (BFGS) algorithm~ \cite{Broyden1970,Fletcher1970,Goldfarb1970,Shanno1970}, as $b_{\rm opt}^{\rm mf} = 0.539$ and $c_{\rm opt}^{\rm mf} = 1.565$ with $\braket{\sigma^z} =1.000$. Notice that no greedy search is applied here since the problem is very simple in the mean-field case. 
Although finding optimal values of the coefficients is sufficient for our purpose in the following analysis, we illustrate in Appendix \ref{appendix:meanfield} the time developments of the wave function, the coefficients, and the magnetization in the mean-field context, comparing mean-field results with the exact solution of the counterdiabatic term, for better understanding of the physics of the mean-field theory.

\subsection{Ferromagnetic system}
\label{subsec:ferro}

\subsubsection{Parameter dependence of the energy and fidelity}

We move a step forward and analyze the ferromagnetic system with all-to-all interactions,
\begin{align}
    \mathcal{H}^z = - \frac{J}{N-1}\sum_{i<j} \sigma_i^z \sigma_j^z. 
\end{align}
The mean-field theory of the previous section is expected to give the correct solution to this model in the thermodynamic limit \cite{nishimori11book} though finite-size effects may reveal differences.

\begin{figure}[thb]
    \centering
    \begin{overpic}[height=32mm]{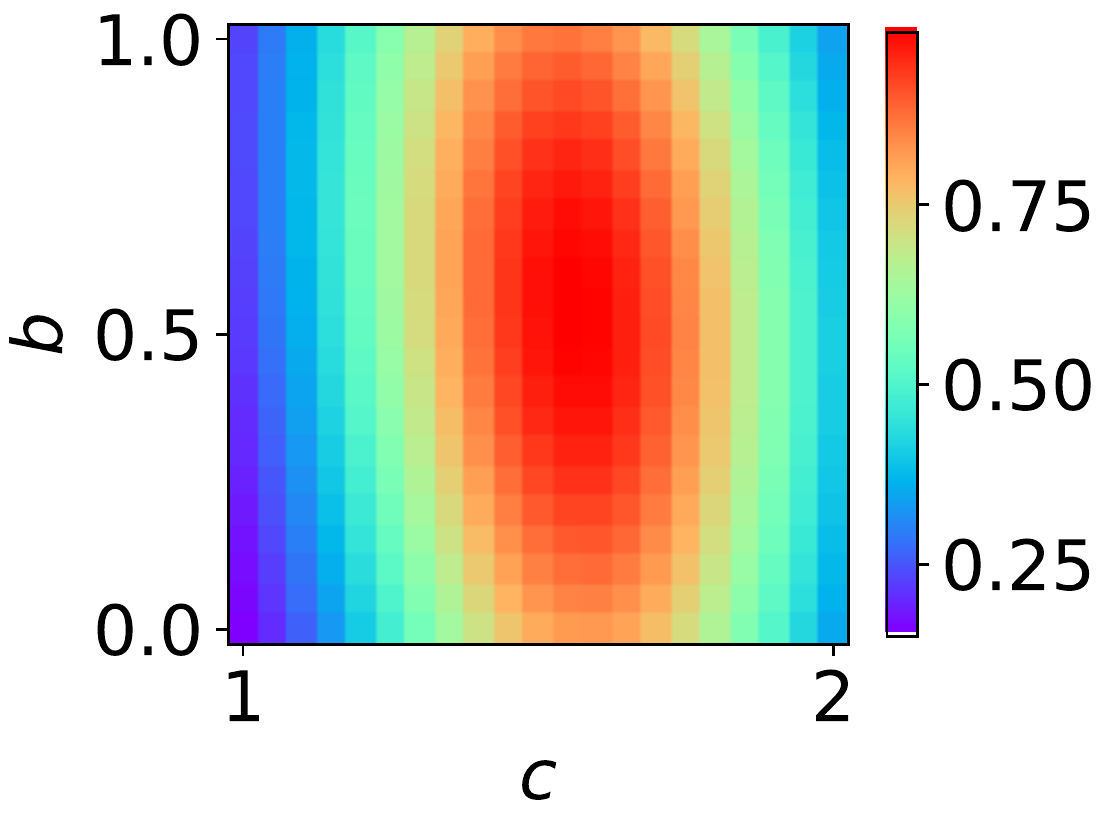}
        \put(0,75){(a)}
    \end{overpic}
    \begin{overpic}[height=32mm]{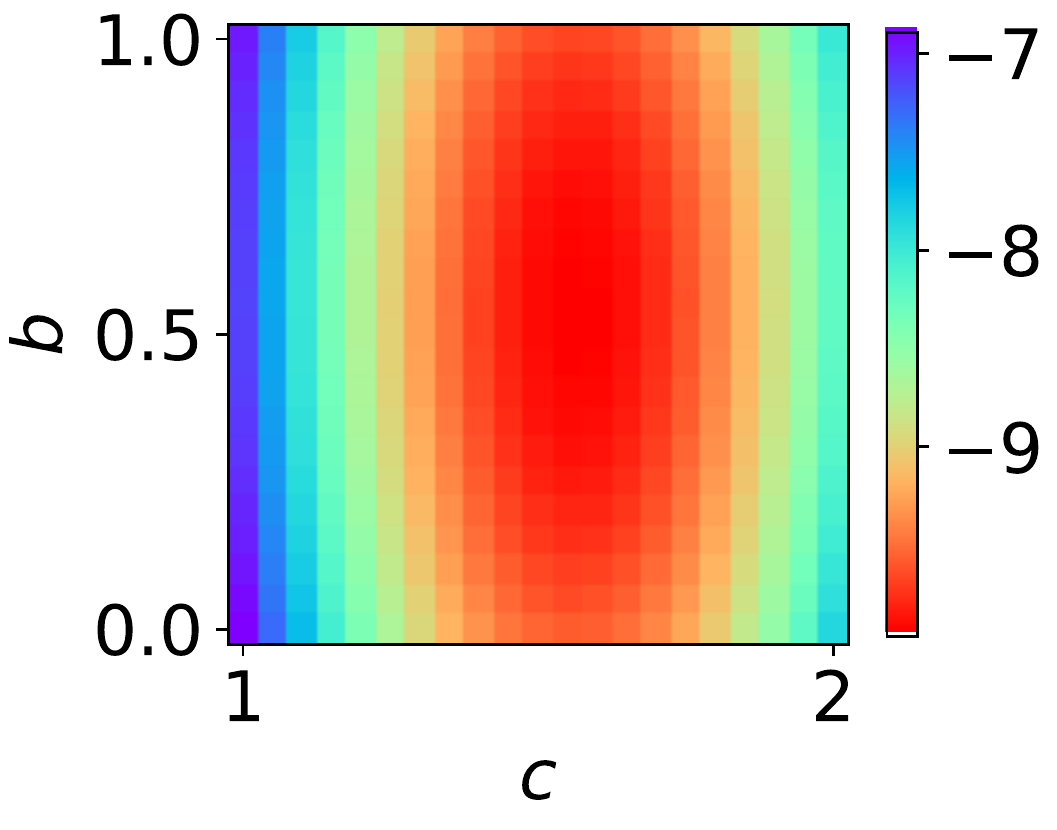}
        \put(0,78.5){(b)}
    \end{overpic}\\
    \begin{overpic}[height=32mm]{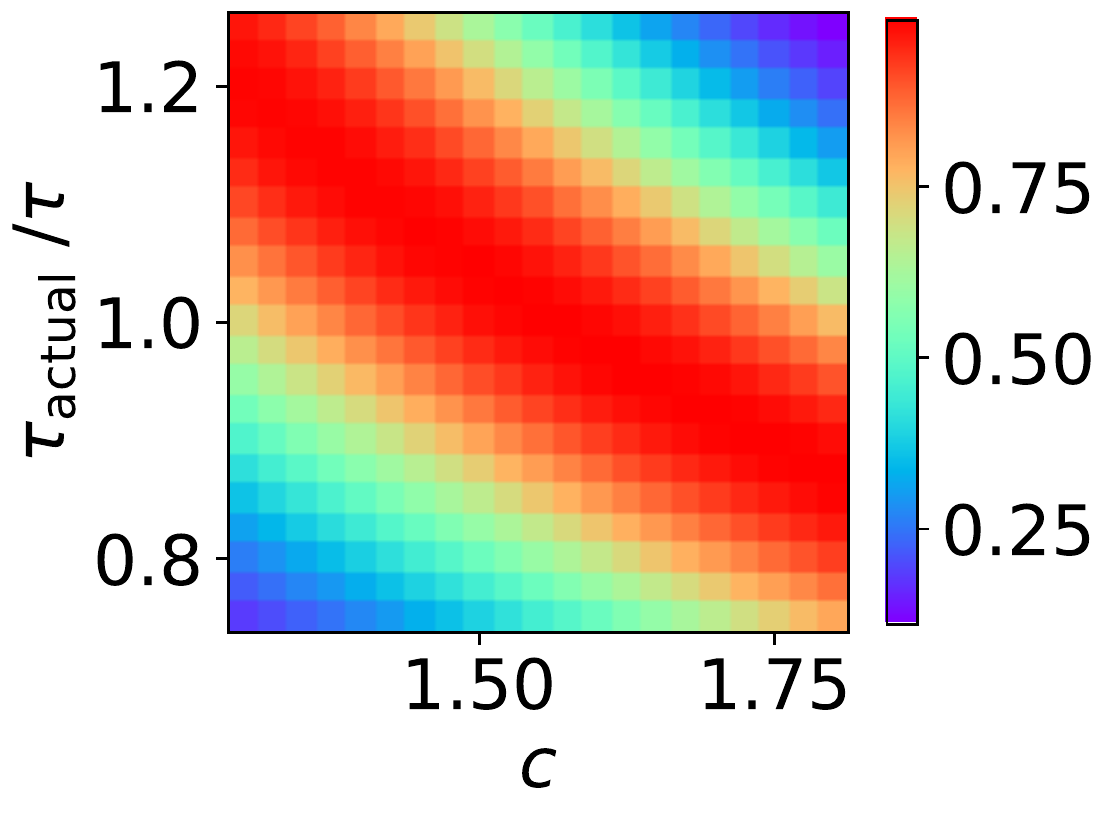}
        \put(0,75){(c)}
    \end{overpic}
    \begin{overpic}[height=32mm]{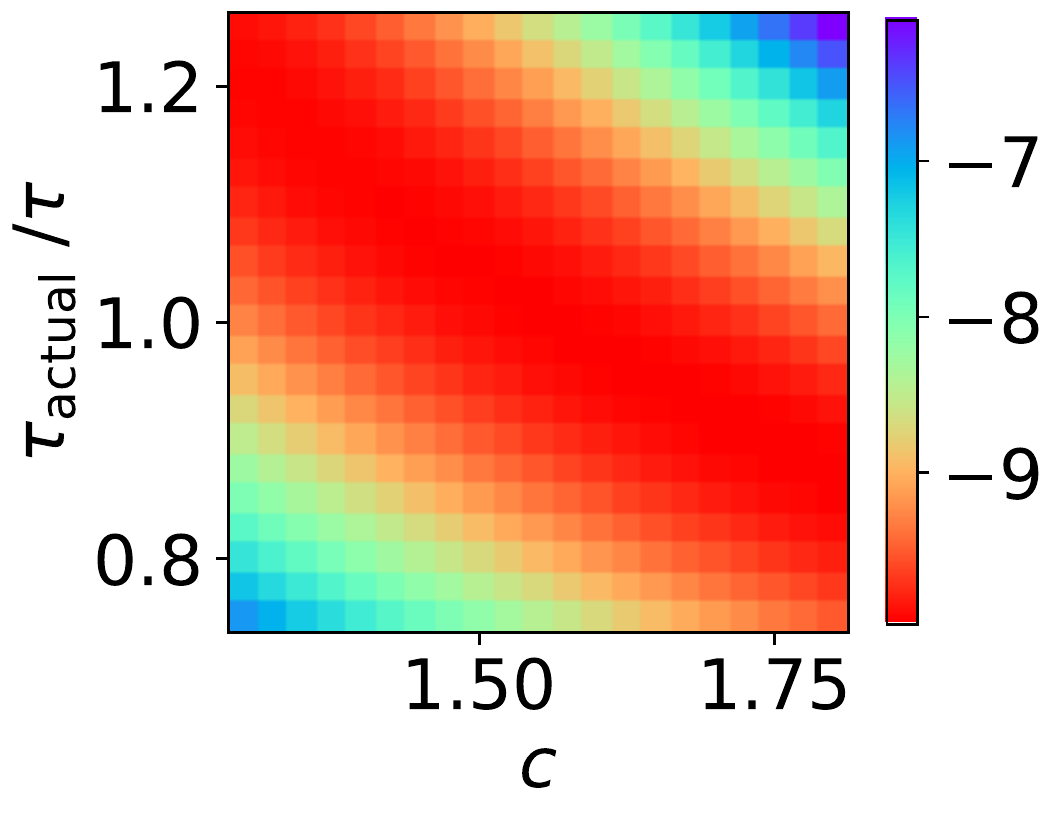}
        \put(0,78.5){(d)}
    \end{overpic}
    \caption{
        \label{fig2}
        Parameter ($b, c$) dependence of (a) the fidelity and (b) the energy in the ferromagnetic model with $N=20$.
        Dependence on the annealing time and the parameter $c$ of (c) the fidelity and (d) the energy when annealing is terminated before the designated annealing time ($\tau_{\rm actual}<\tau$) and is carried through beyond ($\tau_{\rm actual}>\tau$).
        Color scales are located on the right side of each figure.
    }
\end{figure}

As the interactions are homogeneous, we drop the $i$ dependence of $c_i$ and set all of them to $c$. To break the $\mathbb{Z}_2$ symmetry, we choose positive values of $b$ and $c$ and focus our attention on the region around the optimal value obtained in the mean-field case, $b \in [0, 1]$ and $c \in [1, 2]$. Dependence of the fidelity and energy on $b$ and $c$ is shown in Fig.~\ref{fig2}~(a) and (b), respectively, for $N = 20$.
We also conducted the same analysis for the case where annealing is terminated before the designated annealing time ($\tau_{\rm actual}<\tau$) or carried through beyond ($\tau_{\rm actual}>\tau$), and the results are displayed in Fig.~\ref{fig2}~(c) and (d).
These plots indicate that the optimization is robust in $b$ and sensitive to $c$ and the actual annealing time.
Similarities of Fig.~\ref{fig2}~(a) and (b) and of Fig.~\ref{fig2}~(c) and (d) suggest that the two measures of success, the fidelity and the energy, give essentially the same result, as anticipated from the mean-field case.

\begin{figure}[thb]
    \centering
    \begin{overpic}[height=40mm]{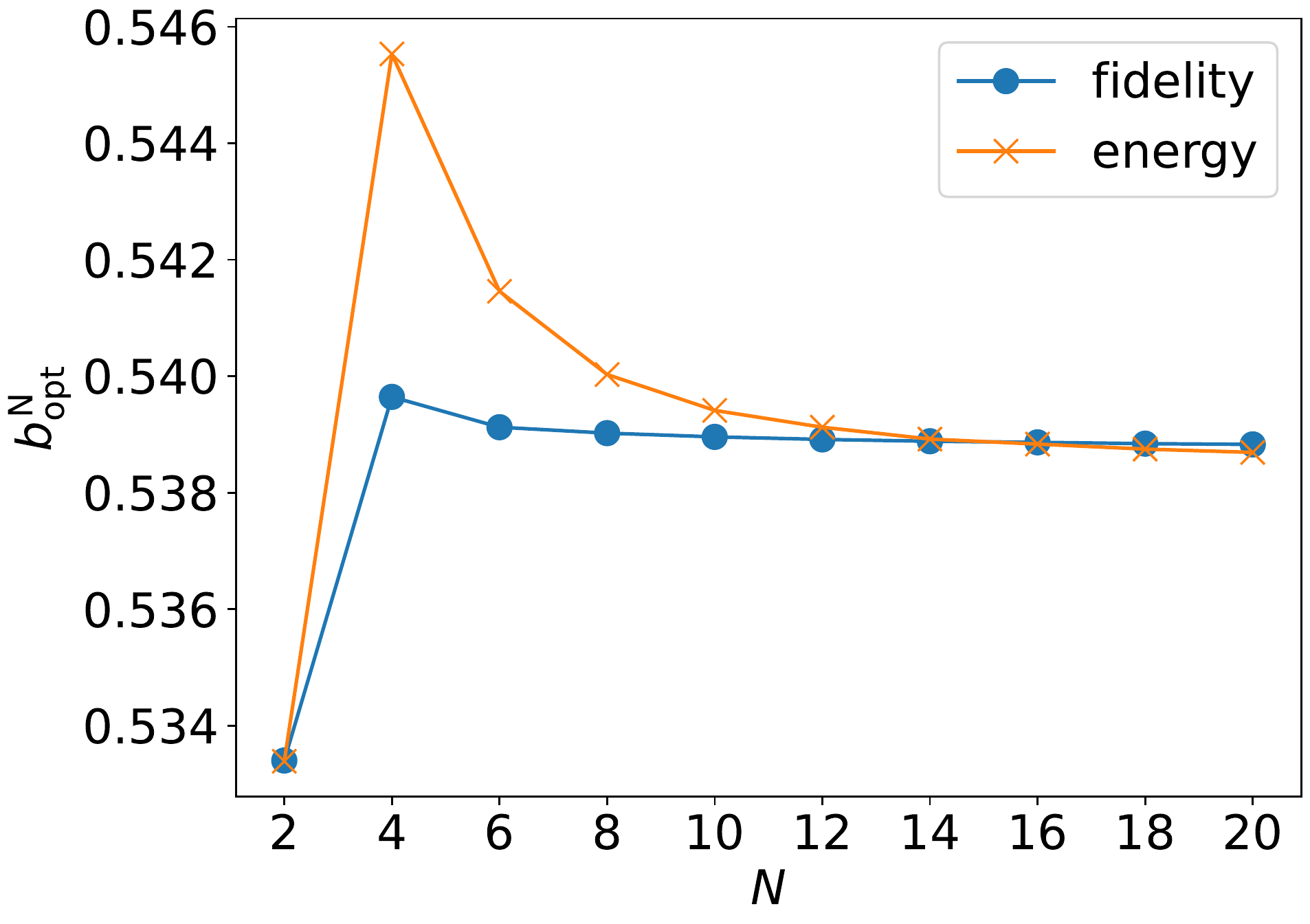}
        \put(0,70){(a)}
    \end{overpic}
    \begin{overpic}[height=40mm]{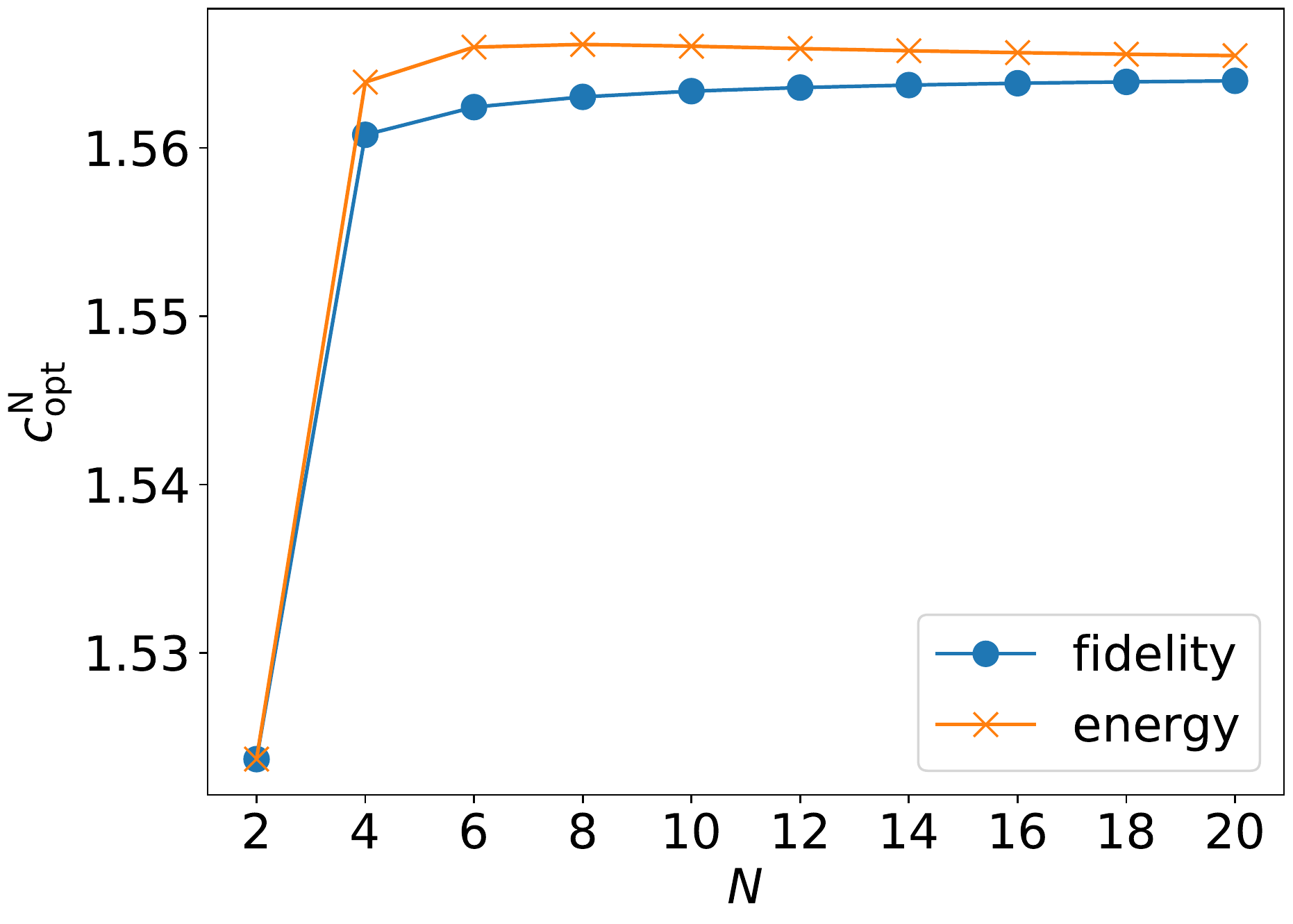}
        \put(0,70){(b)}
    \end{overpic}\\
    \begin{overpic}[height=40mm]{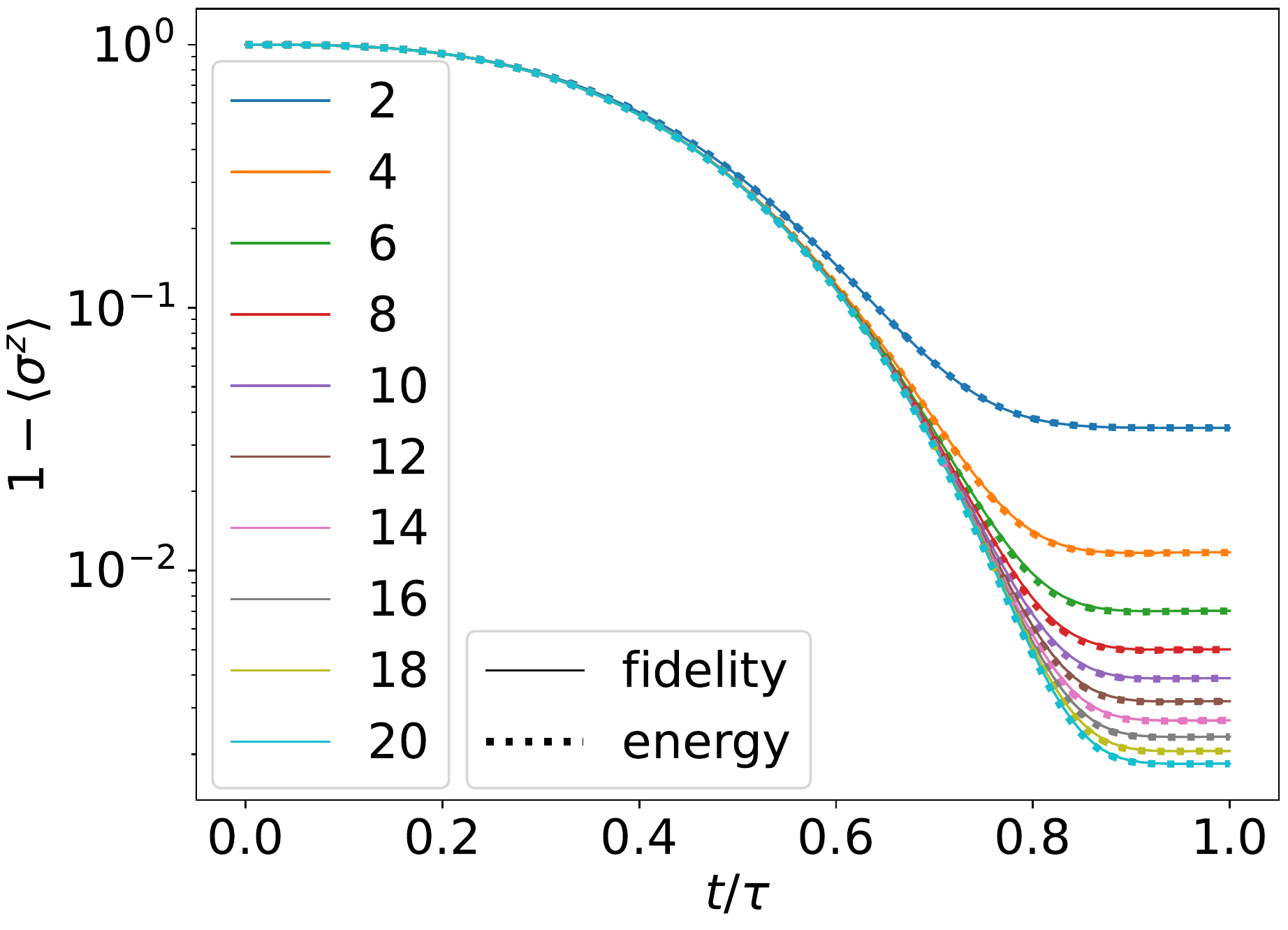}
        \put(0,70){(c)}
    \end{overpic}
    \begin{overpic}[height=40mm]{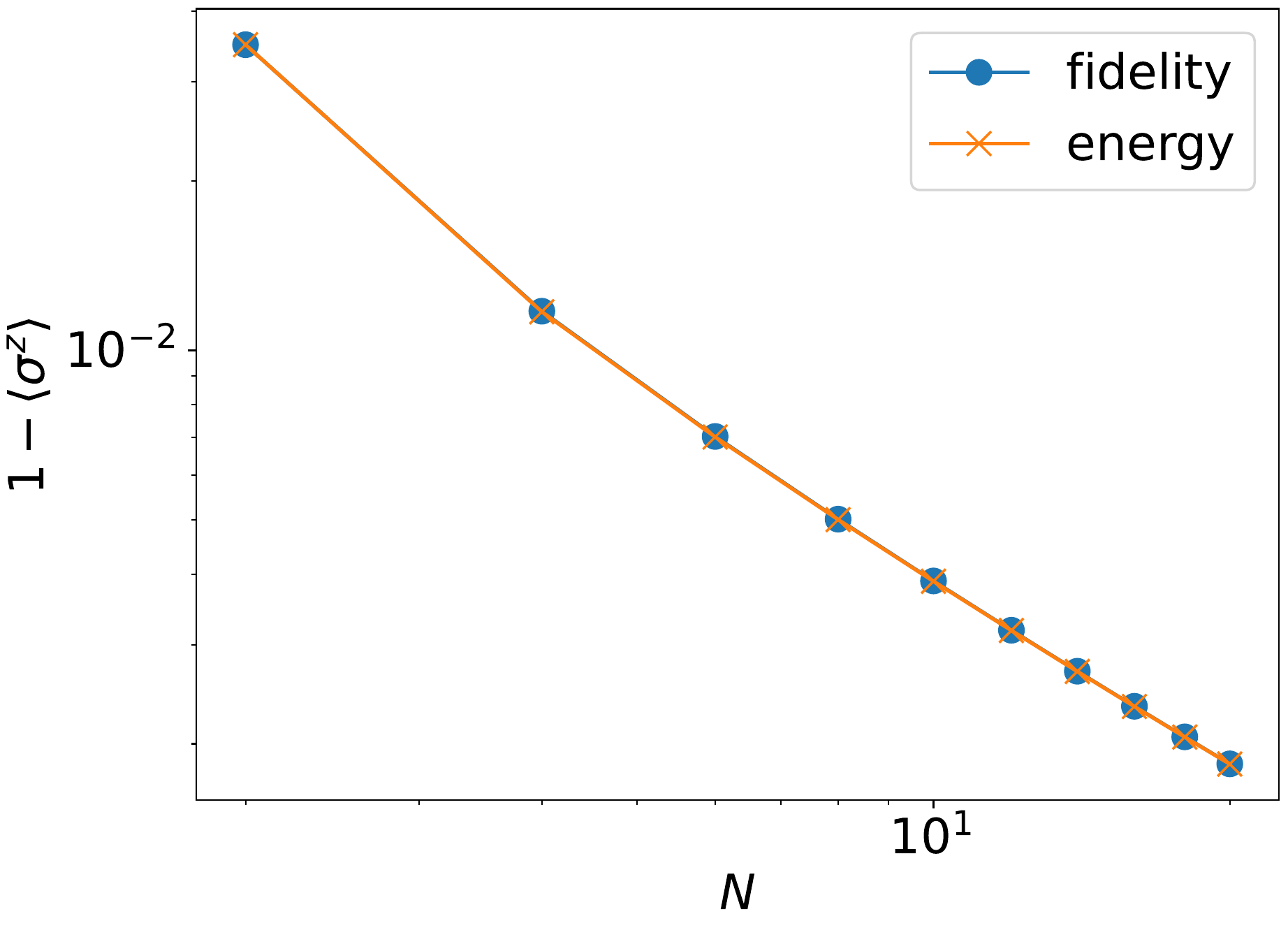}
        \put(0,70){(d)}
    \end{overpic}
    \caption{
        \label{fig3}
        (a) Size dependence of the parameter $b_{\rm opt}^N$  optimized with respect to the fidelity and energy measures in the ferromagnetic model.
        (b) Size dependence of the optimized parameter $c_{\rm opt}^N$.
        (c) Time development of the magnetization, and
        (d) size dependence of the magnetization.
    }
\end{figure}

System size dependence of the optimal parameter values for each $N$, $b_{\rm opt}^N$ and $c_{\rm opt}^N$, is shown in Fig.~\ref{fig3}(a) and (b).
They show convergence beyond $N\approx 12$ toward the values at $N=20$ ($b_{\rm opt}^{N=20} = 0.539, c_{\rm opt}^{N=20} = 1.564$). These values are in good agreement with the results from the mean-field theory, $b_{\rm opt}^{\rm mf} = 0.539$ and $c_{\rm opt}^{\rm mf} = 1.565$, as expected.
The error in magnetization $1-\braket{\sigma^z}$ decreases with the system size, and we do not see a visible difference between the two measures in Fig.~\ref{fig3}~(c) and (d).
The residual error decreases polynomially as a function of the system size $N$.
We will use the size-dependent parameters $b_{\rm opt}^N$ and $c_{\rm opt}^N$ obtained from fidelity optimization for further analysis.

\subsubsection{Greedy optimization of the parameters}

In preparation of the random case, we next try a step-wise greedy optimization of parameters assuming site dependence of the sign of $c_i$ as in the well-established methods like quadratic pseudo-boolean optimization (QPBO) and the roof duality algorithm  solutions~\cite{Kolmogorov2004,Boros2002,Boros2006,Rother2007} and a sampling-based algorithm \cite{Karimi2017}.
More precisely, we choose the optimal values of $c_i$ one by one with other coefficients fixed at their constant values. This approach reduces the search space considerably and also makes the energy (or fidelity) landscape simple.

\begin{figure}[thb]
    \centering
    \begin{overpic}[width=70mm]{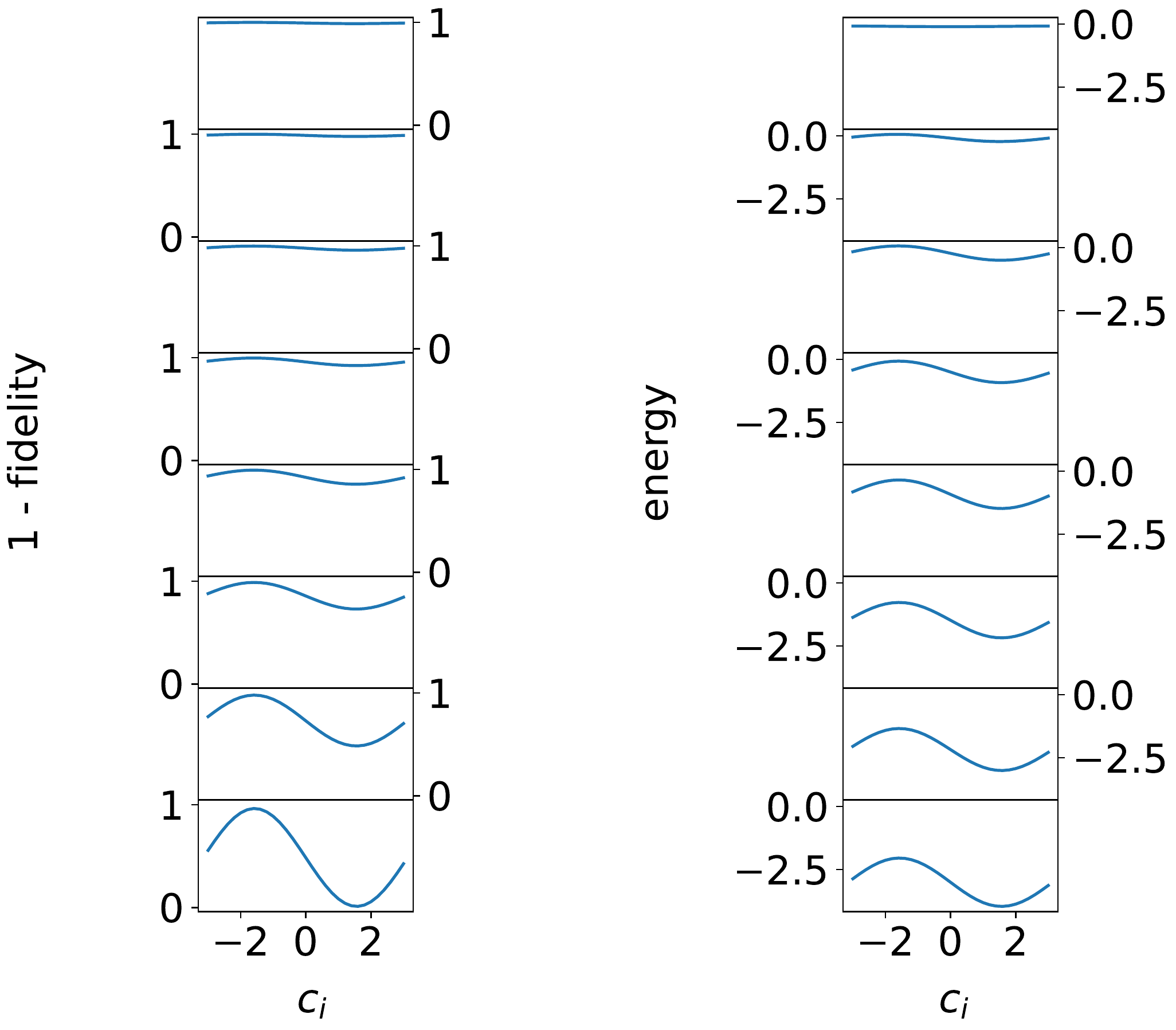}
        \put(1,82){(a)}
        \put(53,82){(b)}
    \end{overpic}
    \caption{
        \label{fig4}
        The parameter $c_i$ dependence of (a) the fidelity and (b) the energy in each iteration (from top to bottom) for the ferromagnetic model with $N=8$. Notice that the site index $i$ changes from step to step as explained in the text. See also Fig.~\ref{figS4} in which the same data are shown but with different vertical scales for different iteration steps for better resolution at each step.
    }
\end{figure}

The first step is to pick up an arbitrary $i$ (e.g., $i=1$) and optimize $c_1$ by setting other $c_j~(j  \ne 1)$ to zero. As shown in the rows of Fig.~\ref{fig4}~(a) and (b) for $N=8$, we choose optimal $c_1$ by minimizing $1-P_{\rm gs}$ (in (a)) or  $E$ (in (b)). See also Fig.~\ref{figS4}, where the scale of the vertical axis is changed step by step for a better resolution. In practice, when we use $E$ as the function to be minimized, this function is symmetric with respect to the change of sign of $c_1$ due to the double degeneracy of states. We therefore arbitrarily set $c_1$ to a positive value $c_1^{\rm BFGS}$ by BFGS.
When we use $1-P_{\rm gs}$ for minimization, we choose the all-up state as the target ground state, so no problem of degeneracy exists.
We next choose another $i$ (e.g., $i=2$) and optimize $c_2$ by fixing $c_1$ to the already-optimized value and other $c_j~(j \ne 1,2)$ to zero. The $c_2$ dependence of $1-P_{\rm gs}$ and $E$ is shown in the second row of Fig.~\ref{fig4}. This process is repeated until all parameters are fixed.
It is seen in Fig.~\ref{fig4} that $1-P_{\rm gs}$ and $E$ become smaller as the iteration proceeds.

We have found that the optimal values $c_i^{\rm BFGS}$ are always close to $c_{\rm opt}^N$: The resulting optimal values vary from site to site, ranging from $1.507$ to $1.566$, while $c_{\rm opt}^{N=8}$ is $1.563$ according to the analysis of the previous section.  This means that our sequential greedy optimization strategy finds the ground state with good precision for the ferromagnetic system with all-to-all couplings.

\subsection{Spin glass problem}
\label{subsec:spinglass}

We apply essentially the same procedure as developed above to the prototypical hard optimization problem of spin glass.  

The algorithm proceeds as follows. The value of $b$ is set to $b_{\rm opt}^N$ obtained for the ferromagnetic case, and the absolute value of $c_i$ is fixed to $c_{\rm opt}^N$ as demonstrated in the ferromagnetic case. We only choose the sign of each $c_i$ step by step, site to site, in a greedy way. It will be seen that this simple process leads to significant improvements in performance as compared to the traditional QA and classical SA.

The sign of $c_i$ is chosen by the average gradient of the optimization measure ($E$ or $1-P_{\rm gs}$) near the origin $0 \le c_i \le \Delta =0.1$  with the values of other $c_j$'s $(j\ne i)$ fixed. The reason is that the optimization measure has only a single minimum, whose position can be detected near the origin, as a function of a single $c_i$, as will be illustrated later.

More precisely, we first choose an arbitrary $i$ (e.g. $i=8$) and fix all other $c_j$ to zero. The optimization measure $E$ is symmetric with respect to the inversion $c_8 \to -c_8$, and we therefore look instead at the curvature of $E$ in the asymmetric range $0\le c_8 \le \Delta$.  This process is repeated for all $i$ and we choose the site $i$ with the largest absolute average curvature. If this turns out to be $i=8$ with a positive derivative, we set $c_8=-c_{\rm opt}^N$, which breaks the symmetry. We next choose another $i$ (e.g. $i=6$) and see the sign of the average derivative of $E$ near the origin with $c_8$ fixed to $-c_{\rm opt}^N$ and all other $c_i$'s to zero. We choose another $i$ and the same process is repeated for all $i(=1, 2, 3, 4, 5,7)$ and we select the site with the largest value of the absolute average derivative (e.g. $i=6$ with a negative derivative) and assign the sign as $c_6=+c_{\rm opt}^N$.  Iterating this process, all $c_i$'s are assigned fixed values.  See Algorithm \ref{alg1}.
\begin{algorithm}[H]
    \caption{Sequential QGO}
    \label{alg1}
    \begin{algorithmic}[1] 
    \REQUIRE QA measure $f(b, {\bm c})$, $b_{\rm opt}^N$, $c_{\rm opt}^N$
    \ENSURE a solution of the cost function
    \STATE $b \leftarrow b_{\rm opt}^N$
    \STATE ${\bm c} \leftarrow (0, \cdots, 0)$
    \REPEAT
    \STATE ${\bm g} \leftarrow \nabla f(b, {\bm c}) = (\frac{f(b, c_1+\Delta, \cdots, c_N) - f(b, {\bm c})}{\Delta}, \cdots, \frac{f(b, c_1, \cdots, c_N+\Delta) - f(b, {\bm c})}{\Delta})$
    \STATE $i \leftarrow \argmax_{j \in \{j|c_j=0\}} | g_j |$
    \STATE $c_i \leftarrow - c_{\rm opt}^N \sgn g_i$
    \UNTIL $c_i \ne 0$ for all $i$
    \RETURN $\sgn {\bm c}$
    \end{algorithmic}
\end{algorithm}

\begin{figure*}[thb]
    \centering
    \includegraphics[height=70mm]{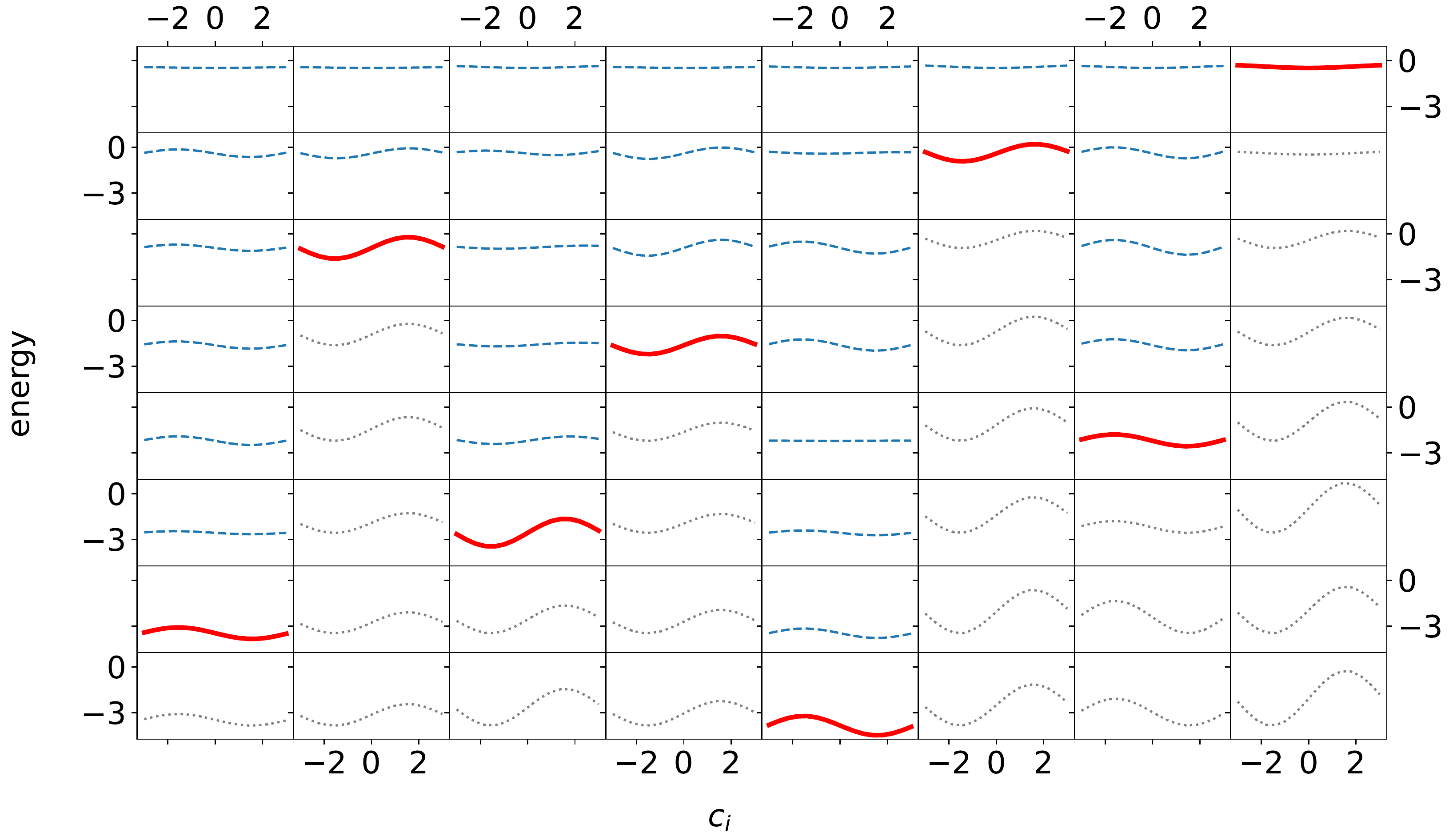}
    \caption{
        \label{fig5}
        Parameter dependence of the energy measure in each iteration of sequential QGO in the random model ($N=8$) from the first step (top row) to the last step (bottom row). Each column represents a site $i$, 1 (leftmost column) to 8 (rightmost column).
        Blue-dashed and red-bold curves are parameters for sites that are under evaluation in each iteration.
        In each iteration, the site with the largest absolute average gradient marked in red-bold is selected to fix the parameter.
        Gray-dotted curves located under the red curves are additional plots for the site which we do not evaluate for further iterations.
        See also Fig.~\ref{figS5} in which the same data are shown with different vertical scales for different iteration steps for a better resolution.
    }
\end{figure*}

An example of this process is illustrated in Fig.~\ref{fig5} for the all-to-all coupling Sherrington-Kirkpatrick model of spin glass \cite{Sherrington1975} with the distribution function of interaction,
\begin{equation}
    P(J_{ij}) \sim \frac{1}{\sqrt{2\pi\sigma^2}} \exp \left( - \frac{J_{ij}^2}{2\sigma^2} \right),
\end{equation}
with  $\sigma^2 = 1/(N-1)$.
Red curves show $E$ as a function of a given $c_i$ which has the largest absolute value of the average gradient. Note that the procedure works for the top row for the initial step as the average gradients are calculated in the asymmetric range $[0, \Delta]$ and resulted in all positive even though the curve is symmetric at $c_i = 0$. Blue dotted curves are for other sites with smaller gradient, and light gray curves show the behavior of $E$ when the already-fixed $c_i$ is tentatively changed with other $c_i$'s fixed. One sees that the minimum value of $E$ on  red curves decreases as the greedy optimization proceeds from the top row to the bottom row. 

The final values of $c_i$ thus obtained are compared with those from direct brute-force optimization by the BFGS algorithm,
\begin{align*}
    {\bm c}_{\rm QGO}  &= (1.563, -1.563, -1.563, -1.563, 1.563, -1.563, 1.563, -1.563)\\
    {\bm c}_{\rm BFGS} &= (1.570, -1.564, -1.562, -1.567, 1.569, -1.561, 1.561, -1.563).
\end{align*}
We observe that the difference is minimal, at most 0.4\%, and the signs are correctly reproduced by the greedy algorithm.

\begin{figure}[thb]
    \centering
    \begin{overpic}[height=45mm]{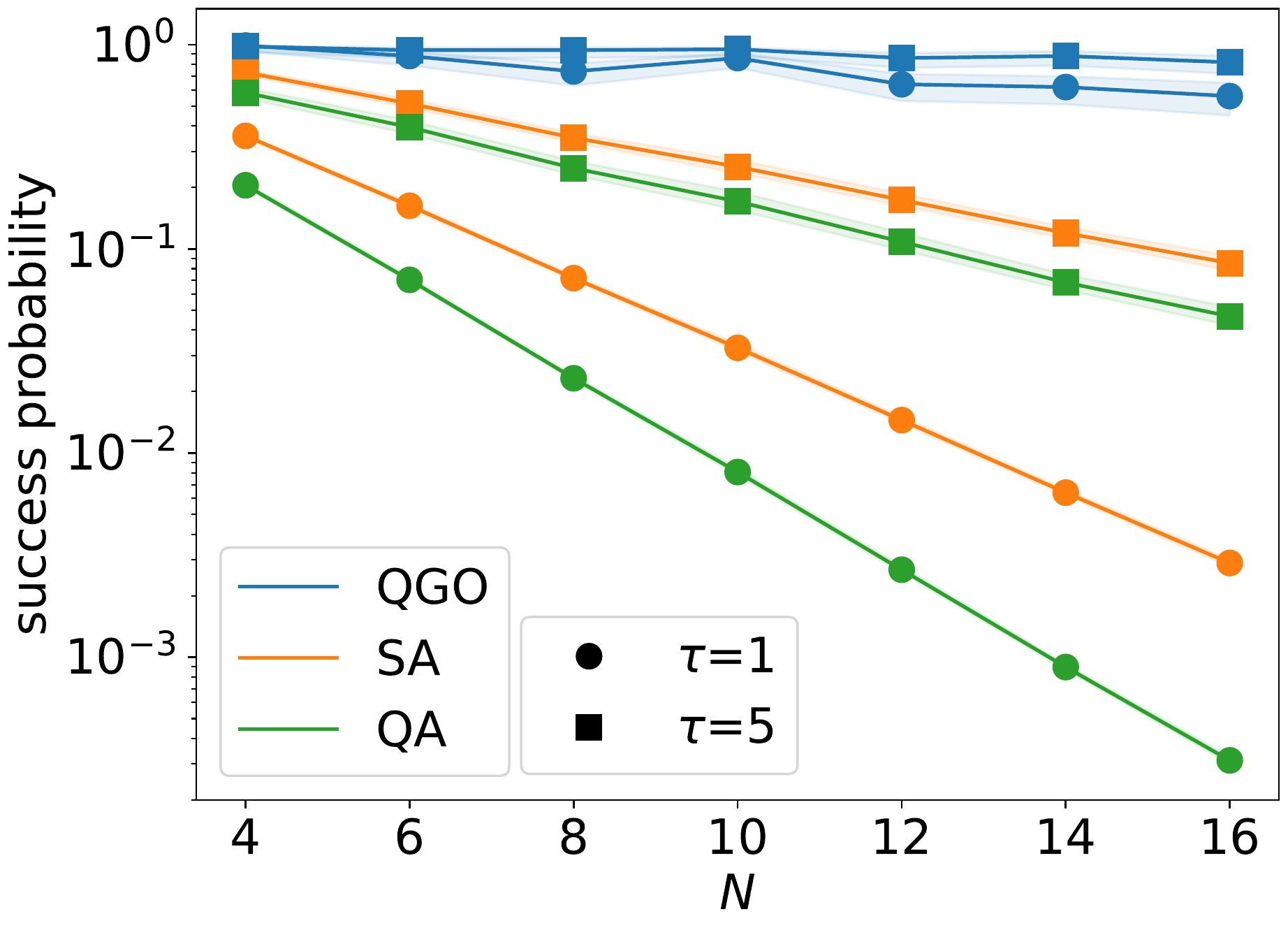}
        \put(0,70){(a)}
    \end{overpic}
    \begin{overpic}[height=45mm]{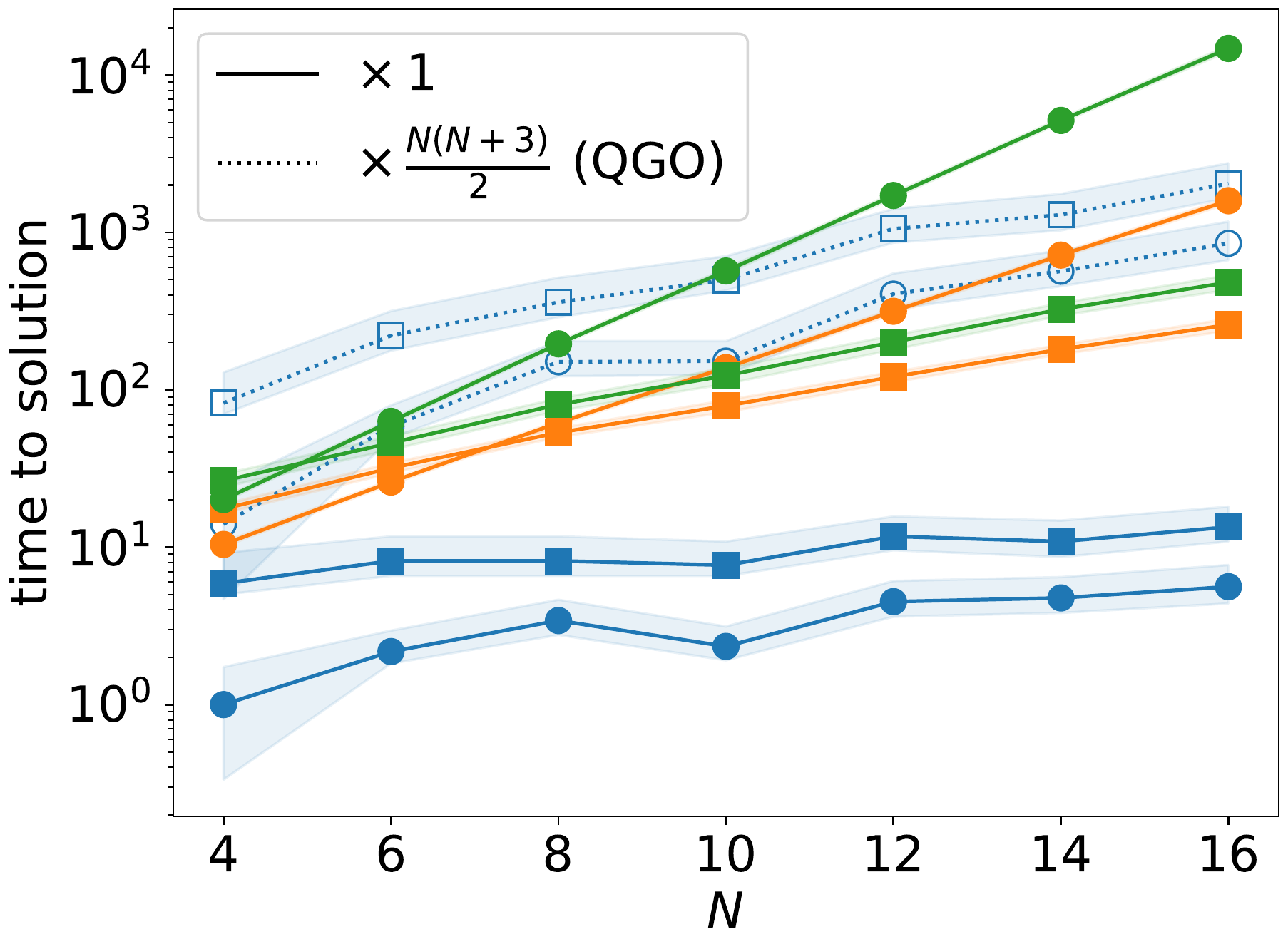}
        \put(0,72){(b)}
    \end{overpic}
    \caption{
        \label{fig6}
        (a) Size dependence of the average success probability of QGO (blue), SA (orange), and QA (green) for $\tau = 1$ (circle) and $5$ (square) in 100 instances of the random system.
        (b) Size dependence of time to solution of QGO (blue), SA (orange), and QA (green) for $\tau = 1$ (circle) and $5$ (square).
        Dotted lines with open markers are plotted by taking into account the factor $N (N+3) / 2$ multiplying the original QGO to compensate for multiple runs of QA.
        The shaded region represents 95\% confidence intervals for each line.
    }
\end{figure}

For a more systematic analysis, we applied the algorithm to the spin glass with various sizes from $N=4$ to 16 with 100 random instances for each $N$.  The resulting success probability is plotted as a function of size in Fig.~\ref{fig6}(a) for $\tau=1$ and 5.  QGO and QA obey the Schr\"odinger dynamics, and the calculation was conducted by QuTiP~\cite{Johansson2013}.
Also plotted are the success probabilities by the traditional QA and SA.  The latter classical algorithm has been tested by the following schedule of temperature decrease (inverse temperature increase)
\begin{align}
    \beta(t) = \frac{(t/\tau)}{1.1 - (t/\tau)}
\end{align}
where $\beta(0) = 0$ and $\beta(\tau) = 10$ and by solving the classical master equation.
QGO returns not a state vector in QA or a probability distribution in SA but a specific solution of spin configuration, which is the output from the Algorithm \ref{alg1}. Thus the success probability of QGO refers to a portion of instances with success.
Confidence intervals are calculated by bootstrapping from 10,000 resamples.
It is clear that the present algorithm QGO far outperforms the other two.

Another viewpoint is provided by the time-to-solution (TTS), a standard measure of computation time for heuristic algorithms \cite{Albash2018}.  It is defined as
\begin{align}
    {\rm TTS} = \tau \frac{\log(1-P)}{\log(1-p_{\rm success})},
\end{align}
where $p_{\rm success}$ is the empirical probability of success for the computation time $\tau$, and $P$ is the target success probability often set to 0.99.  The result is plotted in Fig.~\ref{fig6}~(b), which apparently shows a low computational cost of the present algorithm QGO drawn in blue. We should notice, however, that QGO carries an overhead of repeated computation of the parameter dependence of $E$ at each iteration.

QGO calls QA as a subroutine $N(N+3)/2$ times in total to determine the gradients.
This number is derived as follows:
In the $n$-th iteration ($n = 1, \dots, N$), we run QA at $(c_1, \cdots, c_N)$ once and at $(c_1, \cdots, c_i + \Delta, \cdots, c_N)$ for $i$-th site $N-(n-1)$ times as we already fixed $n-1$ sites.
Thus, the total number is $\sum_{n=1}^{N} (1+N-n+1) = N(N+3)/2$. After we take into account this overhead, the TTS of QGO becomes comparable with the other two as seen in Fig.~\ref{fig6}(b) in blank blue symbols. It is nevertheless concluded that QGO achieves much better success probability with similar computation times.

\begin{figure}[thb]
    \centering
    \begin{overpic}[height=45mm]{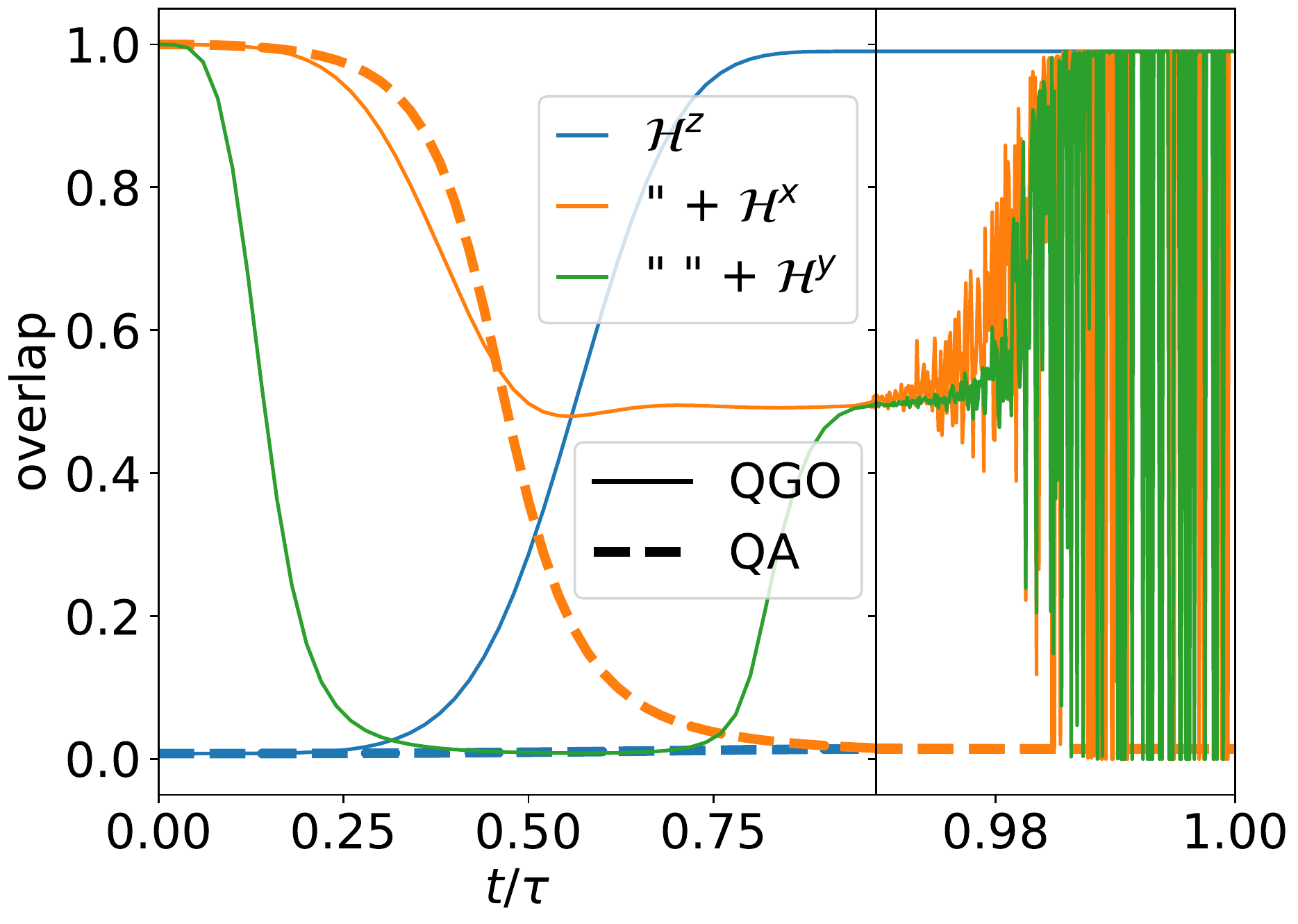}
        \put(0,73){(a)}
    \end{overpic}
    \begin{overpic}[height=45mm]{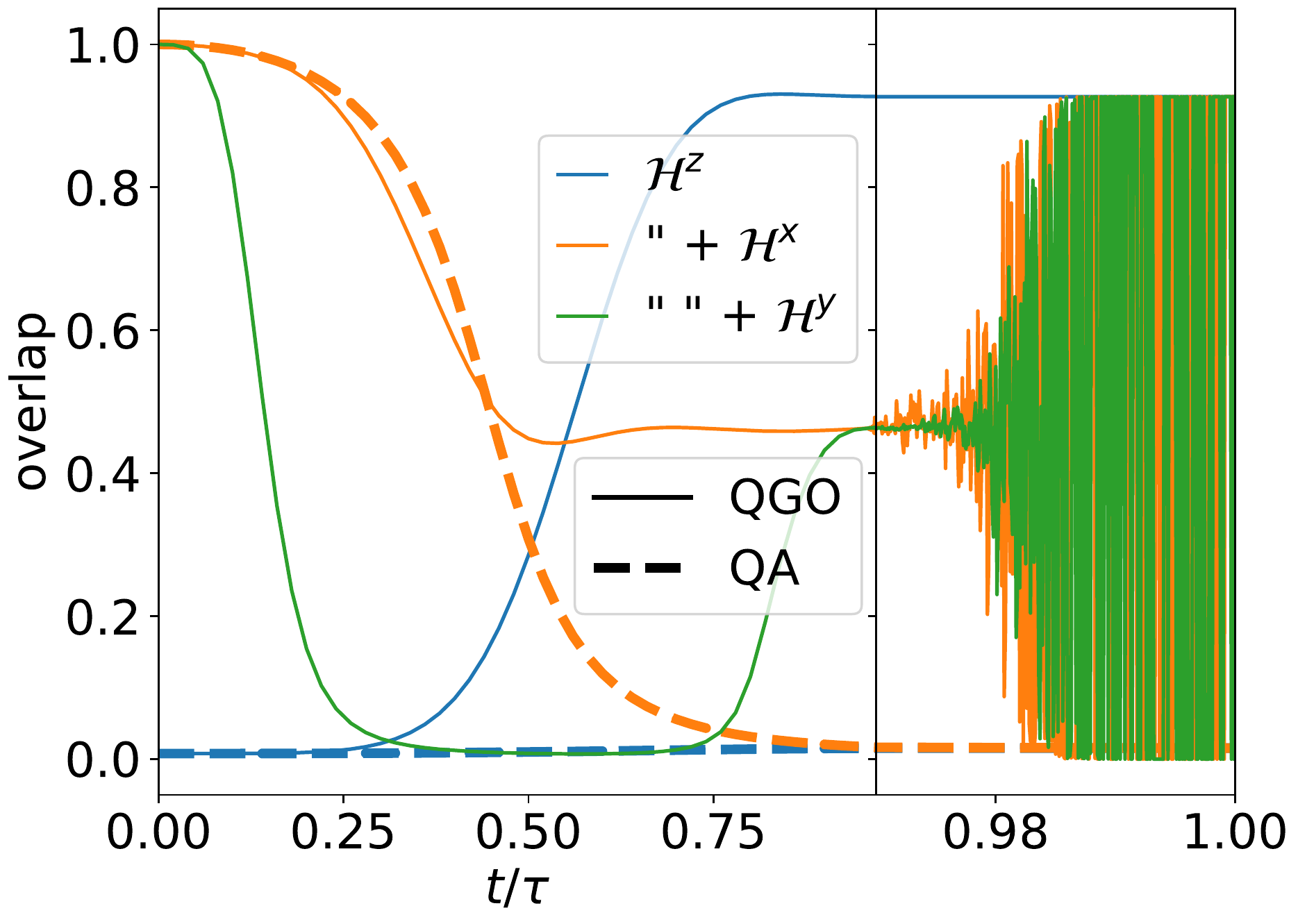}
        \put(0,73){(b)}
    \end{overpic}
    \caption{
        \label{fig10}
        Time dependence of overlaps of the wave function with the instantaneous ground states of three kinds of the Hamiltonian, (blue) the final Ising model $\mathcal{H}^z$, (orange) the transverse-field Ising model $A(s)\mathcal{H}^z+B(s)\mathcal{H}^x$, and (green) the full Hamiltonian $A(s)\mathcal{H}^z+B(s)\mathcal{H}^x+C(s)\mathcal{H}^y$, for (solid) QGO and (dashed) QA.
        The problems are (a) the ferromagnetic system, and (b) a spin glass problem, both with eight spins. Curves in the range [0.97, 100] beyond the black vertical line are magnified for better resolution.
    }
\end{figure}

In order to further understand how the algorithm leads to better results, we have analyzed the time dependence of overlaps of the wave function with the instantaneous ground states of three kinds of the Hamiltonian, (i) the final Ising model $\mathcal{H}^z$, (ii) the transverse-field Ising model $A(s)\mathcal{H}^z+B(s)\mathcal{H}^x$, and (iii) the full Hamiltonian $A(s)\mathcal{H}^z+B(s)\mathcal{H}^x+C(s)\mathcal{H}^y$, for QGO and QA. Results are plotted in Fig.~\ref{fig10}~(a) for the ferromagnetic system and (b) for a spin glass problem.
It is observed that QGO does not directly let the system follow the instantaneous ground state of the full Hamiltonian but it succeeds in enhancing the overlap with the final Ising Hamiltonian at a relatively early stage of computation.

\subsection{Further simplifications of the algorithm}

To see if we can further simplify the algorithm without much compromising the performance, we test two possibilities in this section.

\subsubsection{$y$-field optimization}

\begin{figure}[thb]
    \centering
    \begin{overpic}[width=65mm, clip, trim=80 310 70 320]{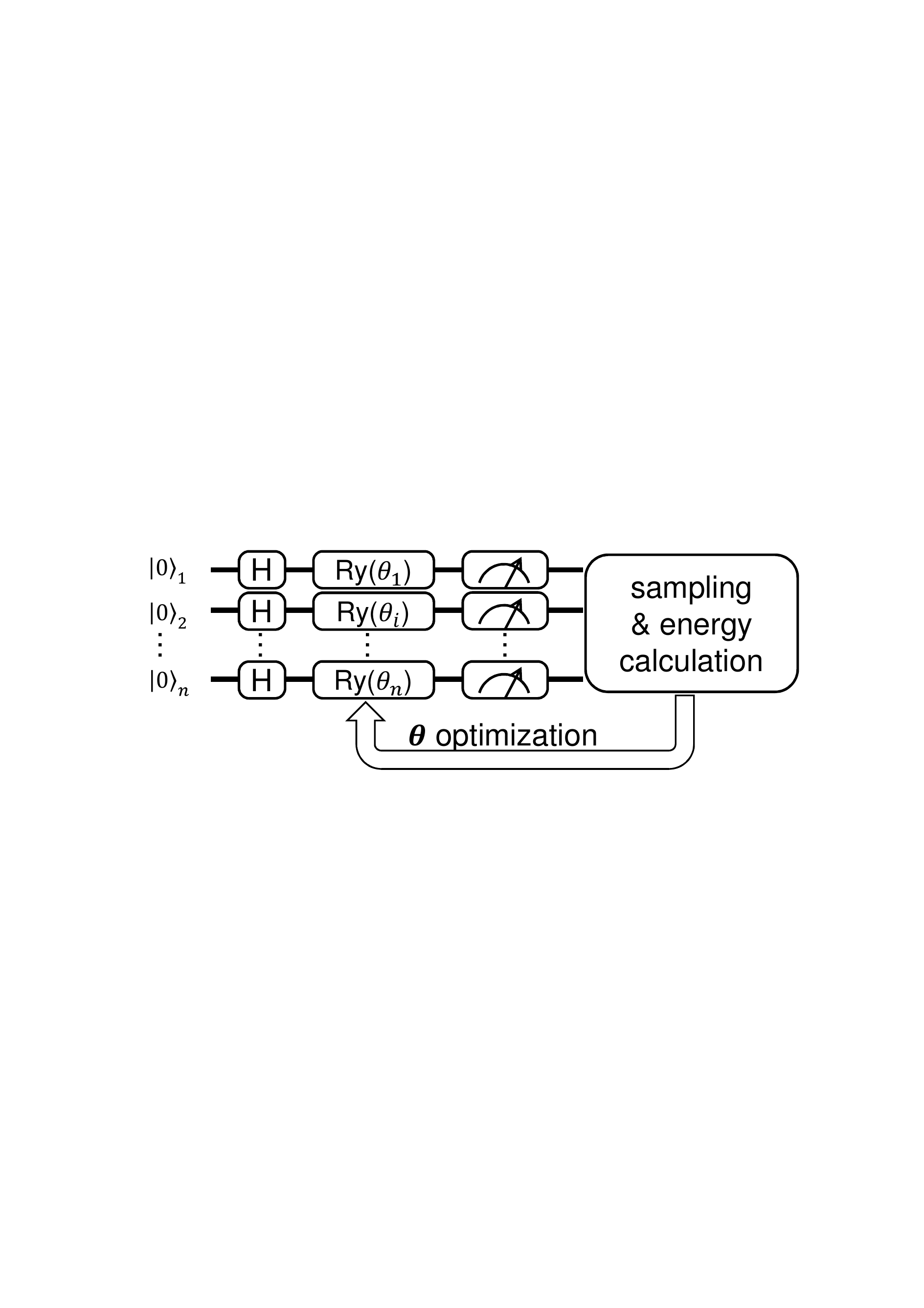}
        \put(-5,32){(a)}
    \end{overpic}
    \begin{overpic}[height=45mm]{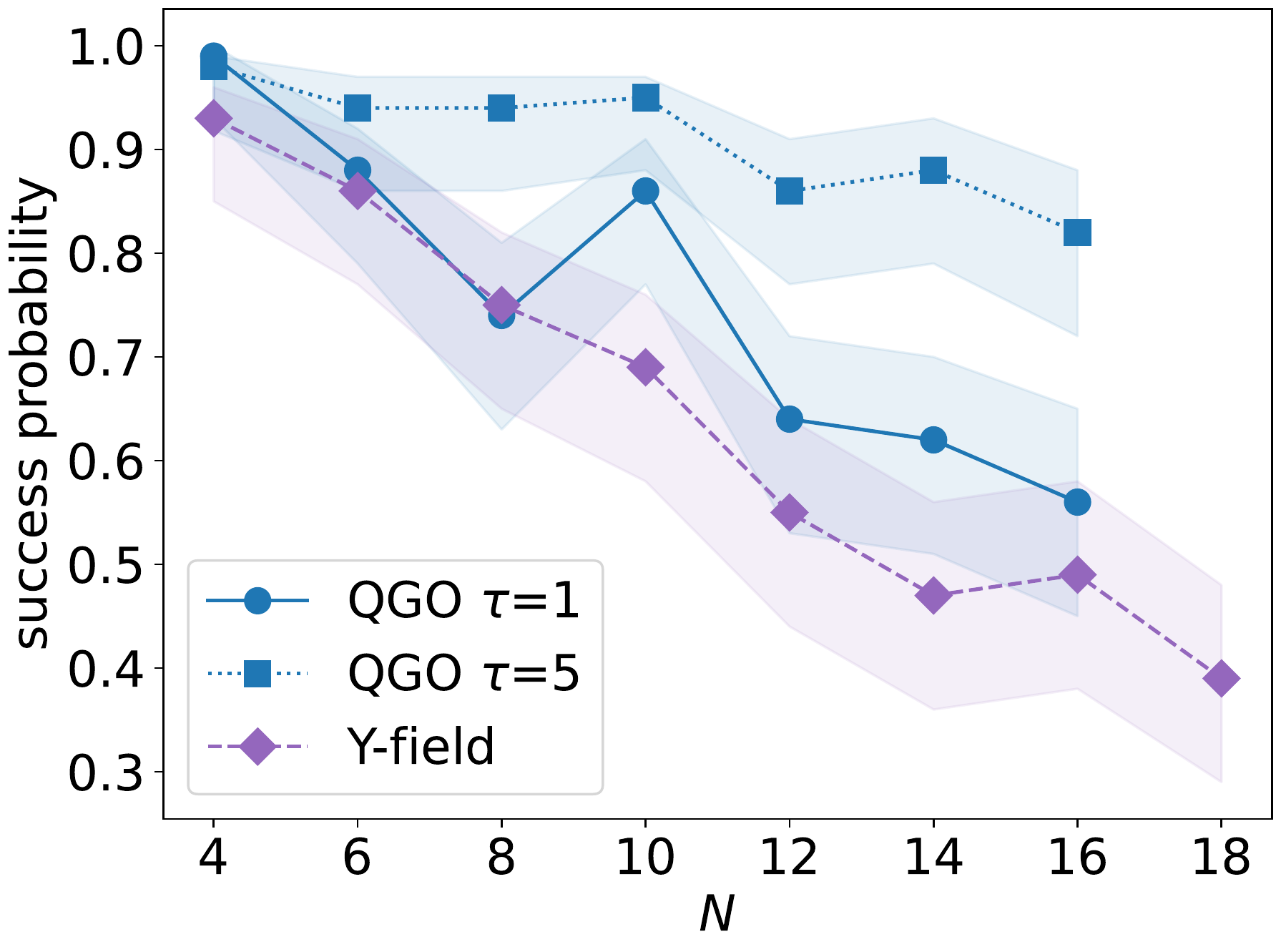}
        \put(0,75){(b)}
    \end{overpic}
    \caption{
        \label{fig7}
        (a) Quantum circuit of the $y$-field optimization.
        (b) Size dependence of the success probability of QGO (blue) for $\tau = 1$ (circle) and $5$ (square) and the $y$-field optimization (purple diamond).
        The shaded region represents 95\% confidence intervals for each line.
    }
\end{figure}

The first example is to drive the state only by $\mathcal{H}^y$, i.e. by a rotation around the $y$ axis,
\begin{align}
    U(\theta) = \exp \left(\frac{i}{2} \sum_i \theta_i(t) \sigma_i^y \right)\equiv \prod_i R_y(\theta_i)
\end{align}
starting from the ground state of $\mathcal{H}^x$ and to variationally optimize the parameters $\theta$ to minimize $E$ as illustrated symbolically in Fig.~\ref{fig7}~(a).
Results for 100 spin glass instances are shown in Fig.~\ref{fig7}~(b).  It is seen that this $y$-field optimization performs slightly worse than QGO for $\tau=1$. It is also noted by comparison with Fig.~\ref{fig6}~(a) that the $y$-field optimization leads to better results than QA and SA with $\tau=1$. Thus the present method may be useful for some purposes due to its simplicity and its easiness of implementation possibly on a gate-based hardware.

\subsubsection{Single-shot QGO}

Let us next study what happens if we fix the signs of coefficients of $c_i$  in a single shot without iteration.  We first fix one of the parameters, e.g. $c_1 = c_{\rm opt}^N$ (spin-up), to break the $\mathbb{Z}_2$ symmetry. Signs of other sites are next fixed according to the gradient of the optimization function, $1-P_{\rm gs}$ or $E$, near the origin as shown in Algorithm \ref{alg2}.
\begin{algorithm}[H]
    \caption{Single-shot QGO}
    \label{alg2}
    \begin{algorithmic}[1] 
    \REQUIRE QA measure $f(b, {\bm c})$, $b_{\rm opt}^N$, $c_{\rm opt}^N$
    \ENSURE a solution of the cost function
    \STATE $b \leftarrow b_{\rm opt}^N$
    \STATE ${\bm c} \leftarrow (c_{\rm opt}^N, 0, \cdots, 0)$
    \STATE ${\bm g} \leftarrow \nabla f(b, {\bm c}) = (\frac{f(b, c_1+\Delta, \cdots, c_N) - f(b, {\bm c})}{\Delta}, \cdots, \frac{f(b, c_1, \cdots, c_N+\Delta) - f(b, {\bm c})}{\Delta})$
    \STATE $c_i \leftarrow - c_{\rm opt}^N \sgn g_i$ for $i \ge 2$
    \RETURN $\sgn  {\bm c}$
    \end{algorithmic}
\end{algorithm}

\begin{figure*}[thb]
    \centering
    \begin{overpic}[width=115mm]{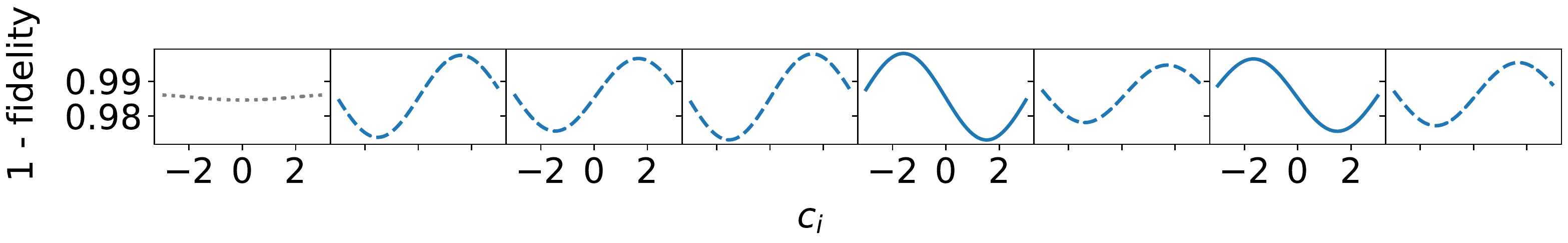}
        \put(3,13){(a)}
    \end{overpic}
    \begin{overpic}[width=115mm]{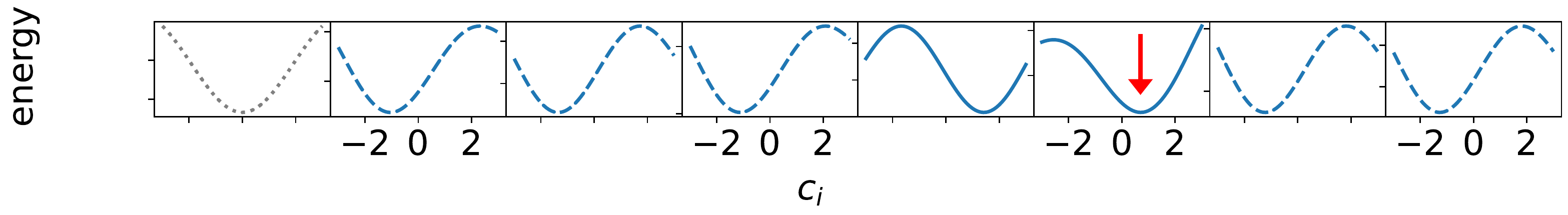}
        \put(3,14){(b)}
    \end{overpic}
    \begin{overpic}[width=115mm]{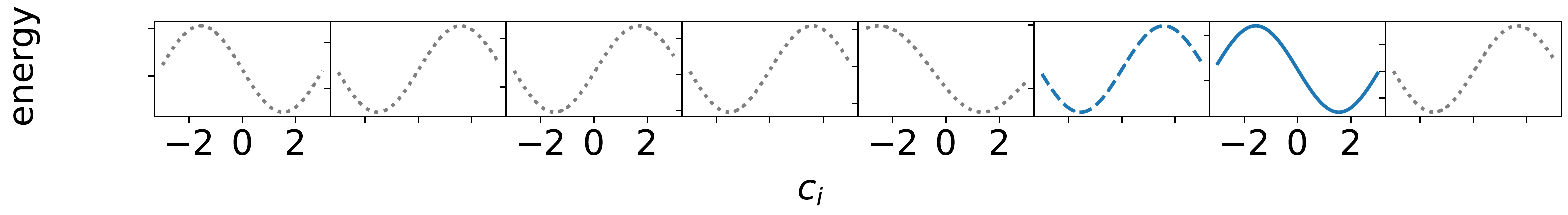}
        \put(3,14){(c)}
    \end{overpic}
    \caption{
        \label{fig8}
        $c_i$ dependence of the single-shot QGO with (a) $1-P_{\rm gs}$ and (b) $E$ for optimization in an eight-spin spin glass problem.
        (c) The sixth and seventh parameters are re-evaluated by fixing other parameters to $\pm c_{\rm opt}$ according to the results in (b).
        Blue-dashed and solid curves represent that the optimal values are negative and positive, respectively.
        Gray-dotted curves indicate re-evaluation of the already fixed parameters similarly to Fig.~\ref{fig5}.
        Vertical scales are normalized for each estimation in (b) and (c).
    }
\end{figure*}

Figure~\ref{fig8}~(a) shows the $c_i$ dependence of $1-P_{\rm gs}$ for each spin to confirm our assumption that the optimal $c_i$ is close to $\pm c_{\rm opt}^N$. Blue-dashed and solid curves represent that the optimal values are negative and positive, respectively. Gray-dotted curves indicate re-evaluation of the fixed-parameter, similarly to Fig.~\ref{fig5}. The correct solution has been obtained by this single-shot QGO using $1-P_{\rm gs}$ for this instance. When the energy measure $E$ is employed for optimization as shown in Fig.~\ref{fig8}~(b), two spins (the sixth and seventh) are fixed incorrectly.
As illustrated in Fig.~\ref{fig8}~(c), the ground state is retrieved if the two spins are re-evaluated after the other six spins are fixed.
This  suggests that we need iterations of the process for better results if we use $E$ for optimization.

It is also noticed that minimum values in the sixth spin in Fig.~\ref{fig8}~(b) are not around $\pm c_{\rm opt}^{N=8} = \pm1.563$ but are around $0.7$ (red arrow).
This implies that our assumption, optimal values are around $\pm c_{\rm opt}^N$ is not valid in this case, which would lead to the difference in the performance between (a) and (b).

\begin{figure}[thb]
    \centering
    \includegraphics[height=45mm]{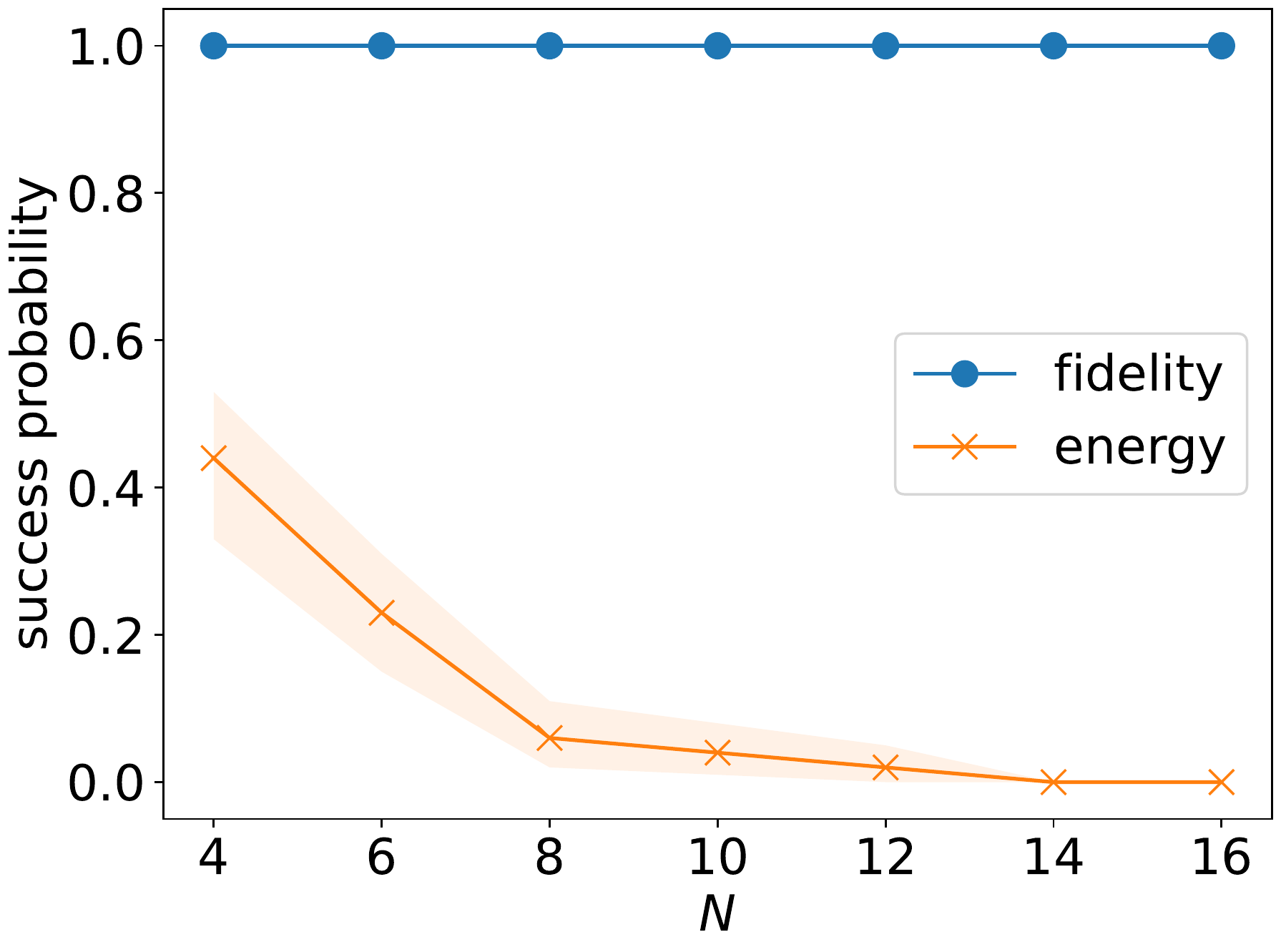}
    \caption{
        \label{fig9}
        Size dependence of the success probability of the single-shot QGO with fidelity (blue circle) and energy (orange x) measures averaged out the 100 instances of spin glass problem.
        The shaded region represents 95\% confidence intervals for each line.
    }
\end{figure}

Figure~\ref{fig9} shows the performances of the single-shot QGO for two measures of optimization for 100 spin glass instances. The single-shot QGO with the fidelity measure always succeeds in identifying the ground state, while the success probability quickly dumps under the energy measure. The fidelity measure is the overlap between the final state and the ground state and thus includes information of the ground state, leading to better results.

\section{Summary and discussion}
\label{sec:summary}

We have formulated a variational algorithm of quantum greedy optimization QGO aiming to achieve improved performance of quantum annealing through a simple process of adjustment of signs of $y$-field coefficients. The result showed notable improvements in the success probability over the original method as well as over classical simulated annealing under comparable computational costs. Since we adjust the signs of coefficients sequentially, the energy landscape of each step is very simple with only a single minimum at almost the same absolute values of coefficients, the adjustment or optimization process  does not encounter the problems of barren plateau in deep variational quantum curcuits \cite{McClean2018} or highly complicated landscapes which plague well-known variational algorithms such as VQE and QAOA \cite{Harrigan2020}. Another advantage of the present method is the simplicity of the additional term in the Hamiltonian, the $y$-field, which can be rewritten in terms of the transverse-field Ising model without $y$ field by a rotation in the spin space. The latter can be implemented experimentally if the hardware can be designed to allow for the non-monotonic time dependence of coefficients of the $x$ field and the Ising part of the Hamiltonian. It is in principle possible to apply the same idea to a final Hamiltonian with $x$ and $y$ components of the Pauli matrix, typically for problems of quantum state preparation in chemistry, although it is non-trivial if the present protocol would lead to satisfactory results in such cases.
In the real settings of quantum processors, we can not measure the fidelity because the target state is unknown. The fidelity analysis provides the algorithm's upper limit, revealing that the success probability can be 100\% and suggesting that the goodness of the cost function defines the algorithm's performance. This finding is informative for the further development of algorithms.

 \enlargethispage{20pt}

\dataccess{The code and data are available in the supplementary materials.} 

\appendix
\counterwithin{figure}{section}
\counterwithin{equation}{section}

\section{Time dependence of the coefficients in the rotated frame}
\label{appendix:coefficients}

\begin{figure}[thb]
    \centering
    \includegraphics[height=45mm]{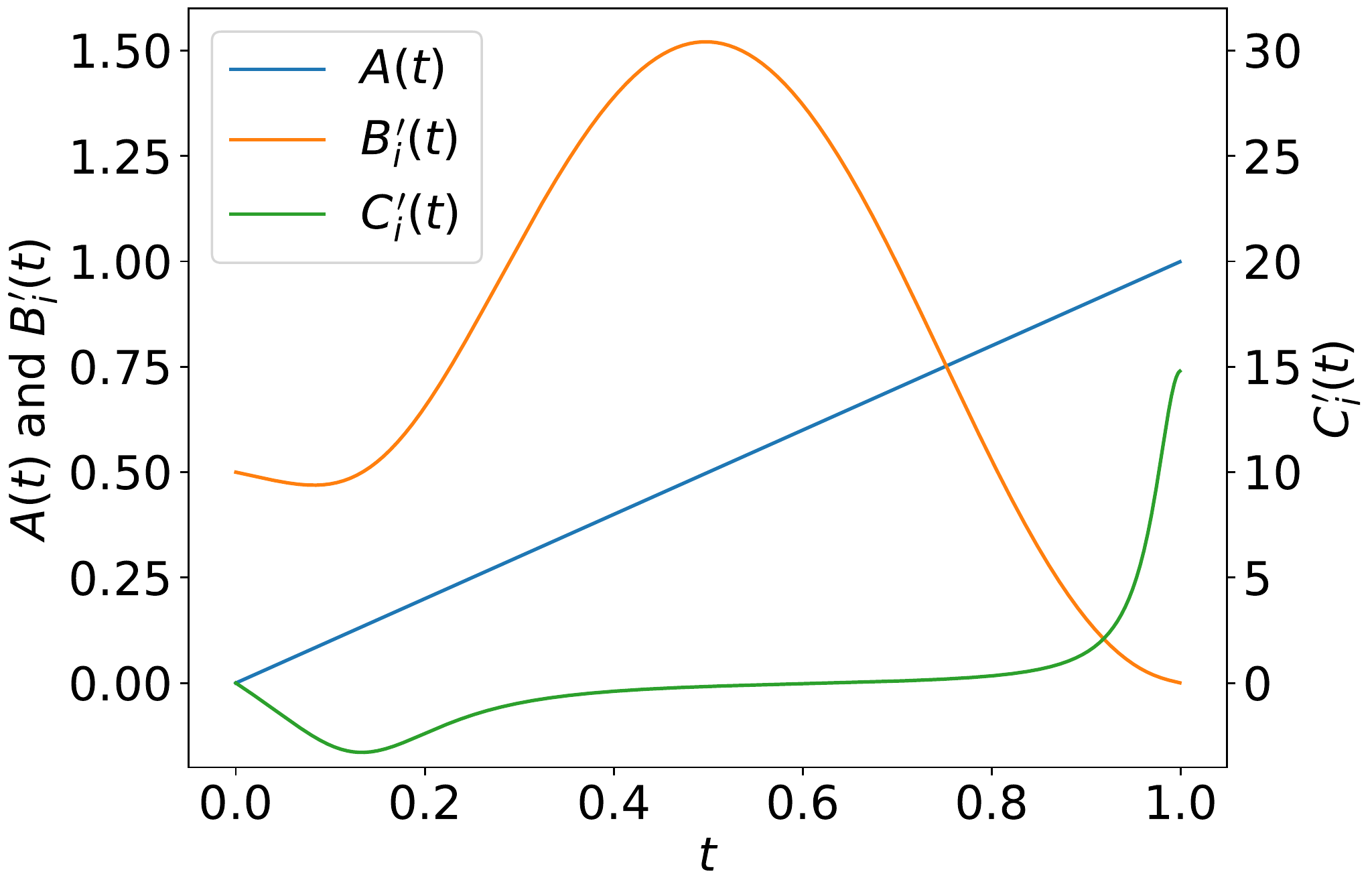}
    \caption{
        \label{figS1}
        Time development of the coefficients $A(t)$, $B_i^\prime(t)$ and $C_i^\prime(t)$  for $a = 1$, $b = 0.5$, $c_i = 1.5$, and $\tau = 1$.
    }
\end{figure}

An example of the behavior of the coefficients in the rotated frame, Eq.~(\ref{eq:Heff}),  is given in this Appendix. Let us write the Hamiltonian as
\begin{equation}
    \mathcal{H}_{\rm eff} =
        A(t) \mathcal{H}^z - \sum_i B_i^\prime(t) \sigma_i^x + \sum_i C_i^\prime(t) \sigma_i^z ,
\end{equation}
where
\begin{eqnarray}
        B_i^\prime(t) & = & \sqrt{B(t)^2 + C_i(t)^2} \\
        C_i^\prime(t) & = & - \frac{bc_i [\pi (1-t/\tau) \sin(2\pi t/\tau) + \sin^2(\pi t/\tau)]}{2\tau [b^2 (1-t/\tau)^2 + c_i^2 \sin^4(\pi t/\tau)]} .
\end{eqnarray}
Figure~\ref{figS1} shows the time dependence of $A(t)$, $B_i^\prime(t)$ and $C_i^\prime(t)$ for $a = 1$, $b = 0.5$, $c_i = 1.5$ and $\tau = 1$.
It is observed that $B'(t)$ and $C'(t)$ are non-monotonic, which suppresses system's transition to excited states in the final state.

\section{System behavior of the mean-field theory}
\label{appendix:meanfield}

\begin{figure}[thb]
    \centering
    \includegraphics[height=65mm]{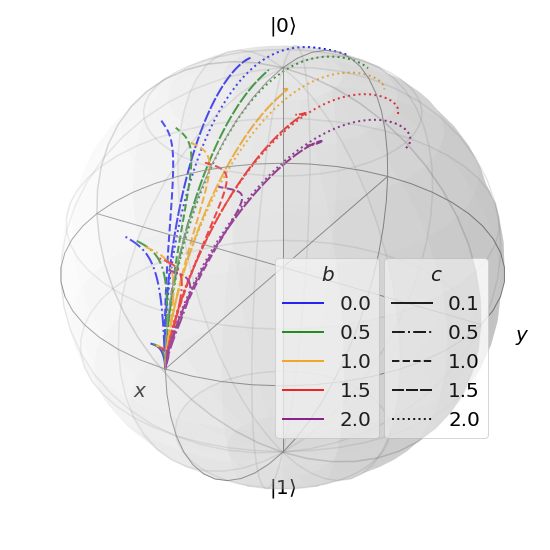}
    \caption{
        \label{figS2}
        Trajectories of the system under the mean-field theory for $b =$ (0.0, 0.5, 1.0, 1.5, 2.0) and $c =$ (0.1, 0.5, 1.0, 1.5, 2.0).
    }
\end{figure}

To visualize the system behavior under the mean-field theory discussed in part~\ref{subsec:meanfield} of section \ref{sec:numerical}, we plot the trajectory of the system on the Bloch sphere in Fig.~\ref{figS2} for a number of parameter values with $b=0.5$ (green) and $c=1.5$ (long dashed) being closest to the optima.
The system starts along the $x$ axis and is supposed to end ideally at the north pole $|0\rangle$. The optimal parameter values are observed to approximately follow a straightforward path connecting these points.

\begin{figure}[thb]
    \centering
    \begin{overpic}[height=45mm]{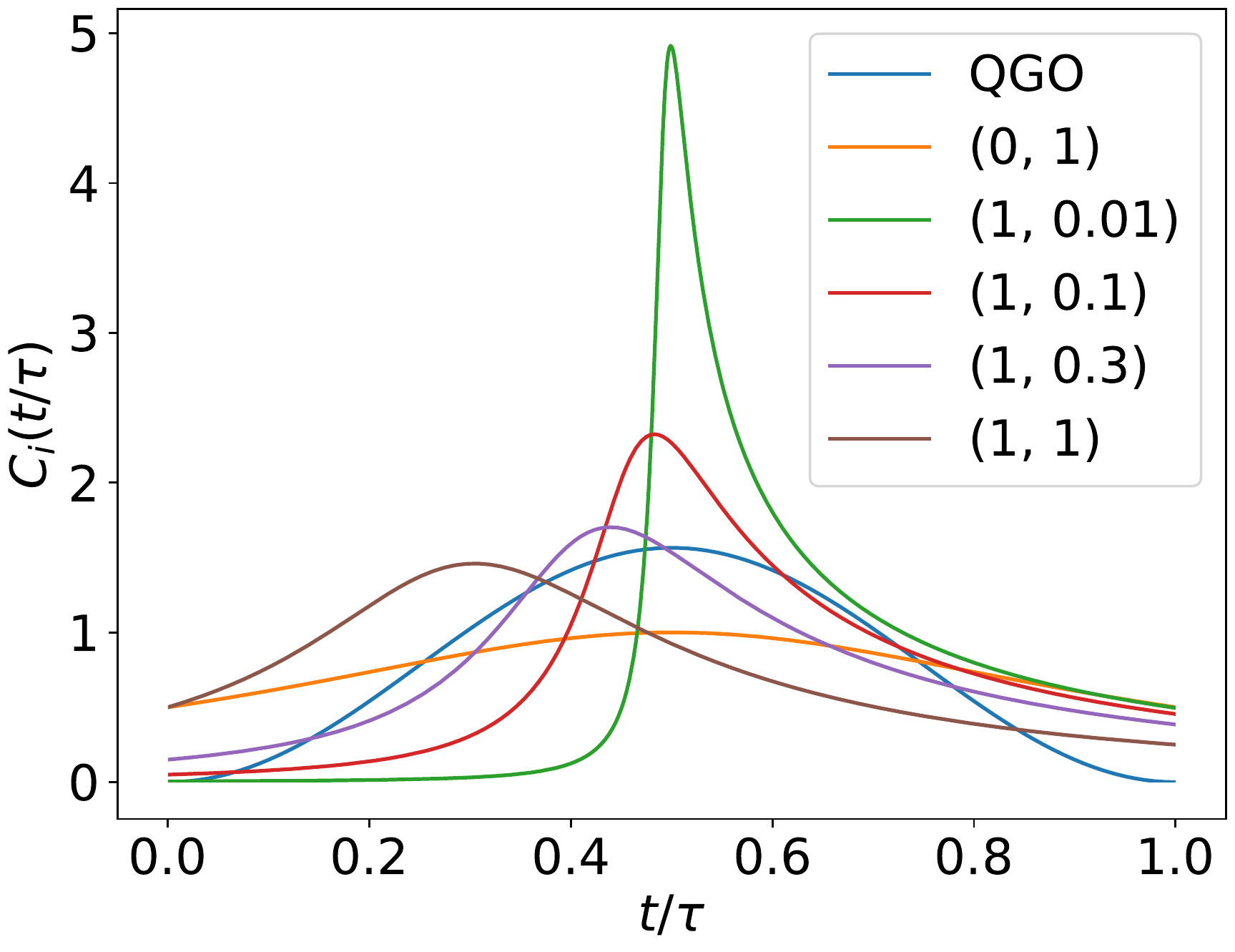}
        \put(0,78){(a)}
    \end{overpic}
    \begin{overpic}[height=45mm]{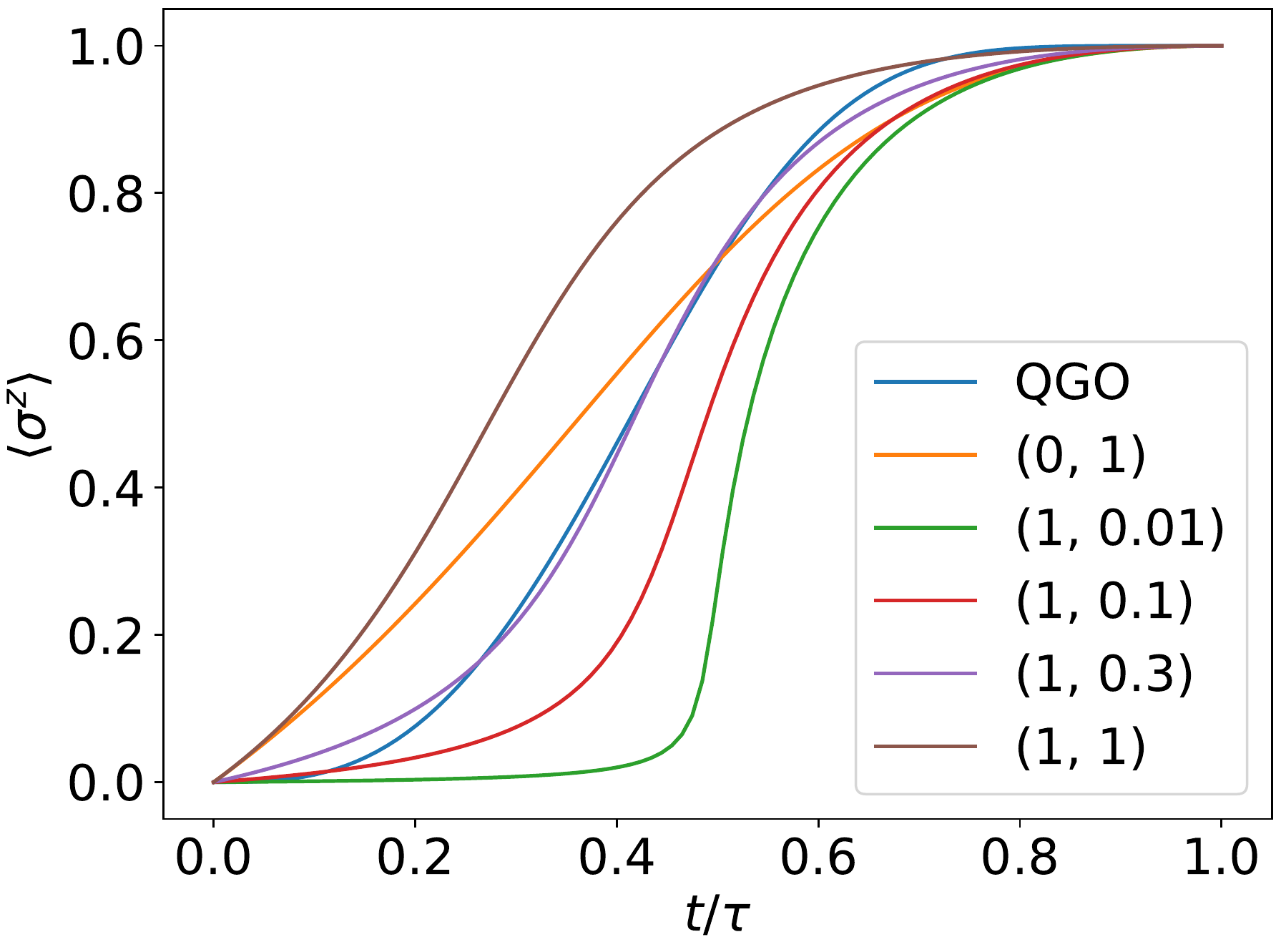}
        \put(3,75){(b)}
    \end{overpic}
    \caption{
        \label{figS3}
        Exact solution of (a) counter-diabatic Y-field, and (b) magnetization for QGO and $(g, h) =$ $(0, 1)$, $(1, 0.01)$, $(1, 0.1)$, $(1, 0.3)$ and $(1, 1)$.
    }
\end{figure}

Figure~\ref{figS3} compares the exact counterdiabatic driving function for the mean-field theory \cite{Takahashi2013} and our approximate result.
The parameterized Ising Hamiltonian is
\begin{equation}
	\mathcal{H}^z = -g \braket{\sigma^z} \sigma^z -h \sigma^z .
\end{equation}
Although QGO tested in the main text is $g = 1$ and $h = 0$, the parameter $h$ should be non-zero in the exact calculation to avoid the double degeneracy of states.
Exact solutions are obtained for $(g, h) =$ $(0, 1)$, $(1, 0.01)$, $(1, 0.1)$, $(1, 0.3)$ and $(1, 1)$.
The system follows the instantaneous ground state under these exact solutions whereas it does not under our approximate scheme, to which $(g, h) = (1, 0.3)$ shows the closest behavior.

\section{Enhanced scales of performance measure}
\label{appendix:enhanced_scale}

\begin{figure}[thb]
    \centering
    \begin{overpic}[width=70mm]{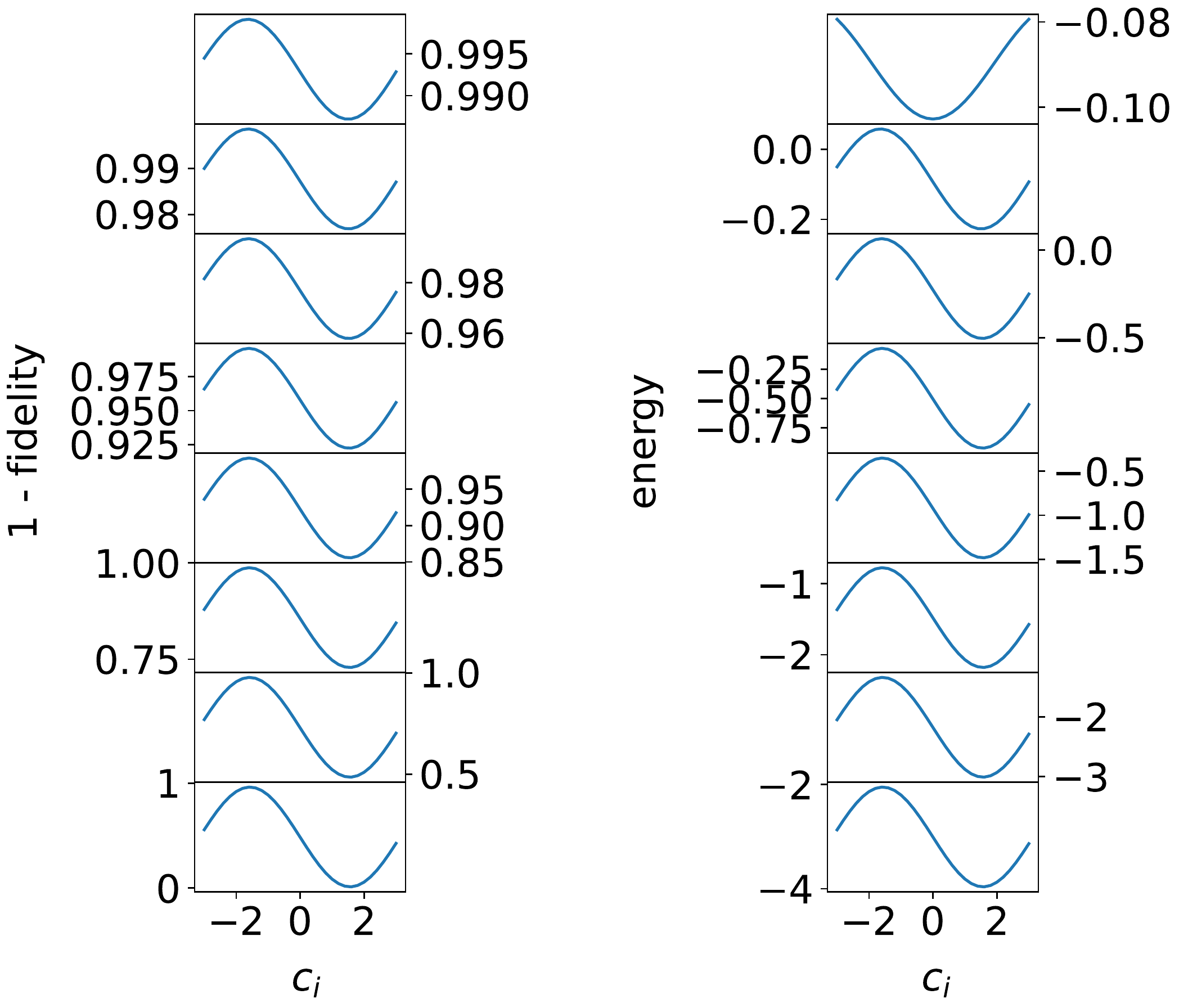}
        \put(1,82){(a)}
        \put(53,82){(b)}
    \end{overpic}
    \caption{
        \label{figS4}
        The same plot as in Fig.~\ref{fig4} with different vertical scales.
    }
\end{figure}

\begin{figure}[thb]
    \centering
    \includegraphics[height=70mm]{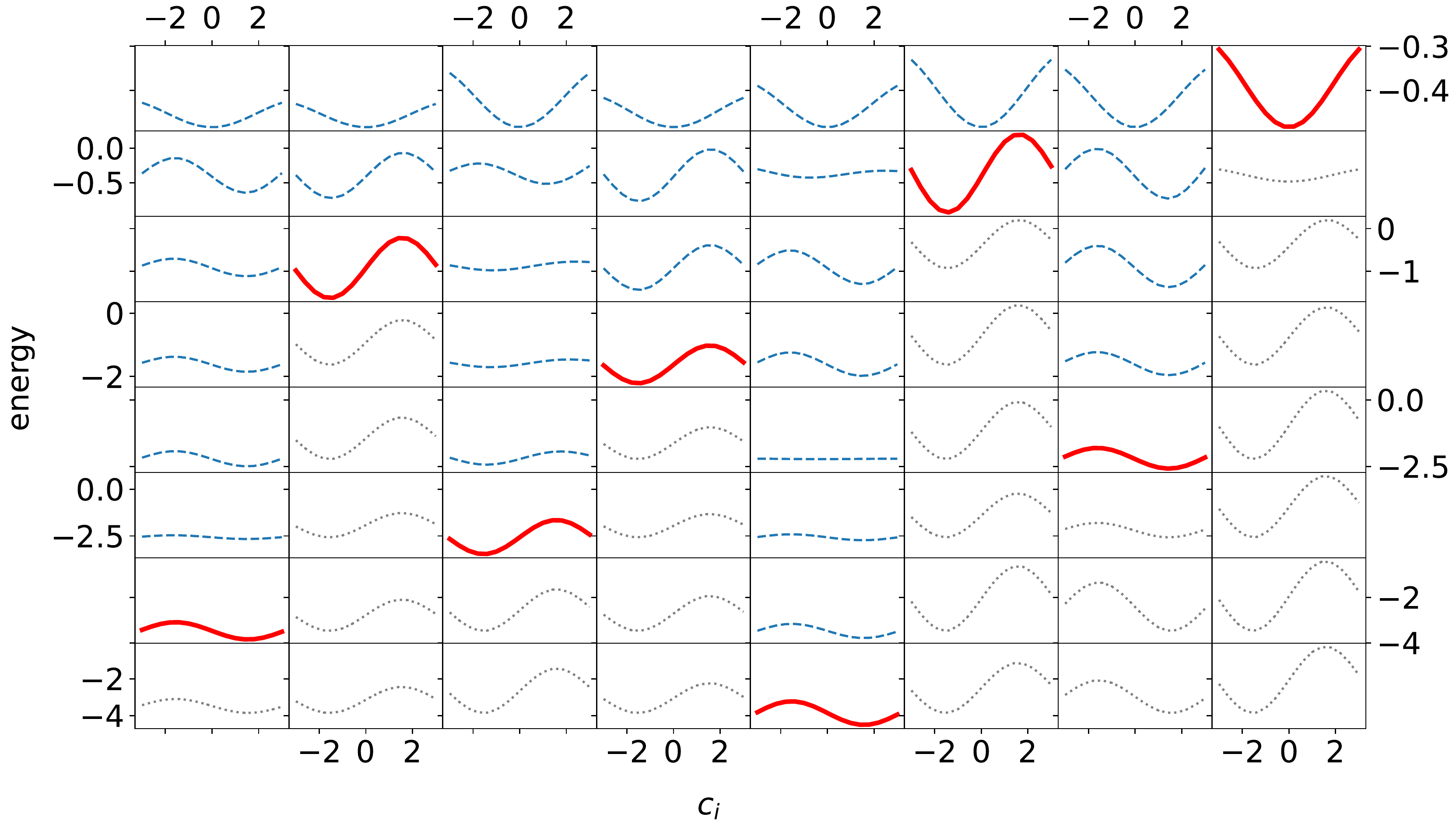}
    \caption{
        \label{figS5}
        Same plot as in Fig.~\ref{fig5} with different scales of vertical axis.
    }
\end{figure}

For better resolution of the vertical axes of Figs.~\ref{fig4} and \ref{fig5}, we show in Figs.~\ref{figS4} and \ref{figS5} the same data with enhanced scales of vertical axes.

\section{$y$-field optimization}

The simple $y$-field optimization algorithm can be executed the circuit model in Fig.~\ref{fig7}(a), and described as follows:
\begin{algorithm}[H]
    \caption{Simple Y-field optimization}
    \label{algS1}
    \begin{algorithmic}[1] 
    \REQUIRE energy measure $f({\bm \theta})$
    \ENSURE a solution of the cost function
    \STATE ${\bm \theta} \leftarrow (0, \cdots, 0)$
    \REPEAT
    \STATE ${\bm g} \leftarrow \nabla f({\bm \theta}) = (\frac{f(\theta_1+\Delta, \cdots, \theta_N) - f({\bm \theta})}{\Delta}, \cdots, \frac{f(\theta_1, \cdots, \theta_N+\Delta) - f({\bm \theta})}{\Delta})$
    \STATE $i \leftarrow \argmax_{j \in \{j|\theta_j=0\}} | g_j |$
    \STATE $\theta_i \leftarrow - \frac{\pi}{2} \sgn g_i$
    \UNTIL $\theta_i \ne 0$ for all $i$
    \RETURN $\sgn {\bm \theta}$
    \end{algorithmic}
\end{algorithm}


\bibliographystyle{unsrtnat}



\end{document}